\documentclass[useAMS,usenatbib,usegraphicx,referee]{mn2e}
\usepackage{natbib}
\usepackage{lscape}
\usepackage{longtable}

\newcommand{\msun}{M_\odot}

\newcommand{\gyr}{\rm{Gyr}}
\newcommand{\dens}{M_\odot \rm{pc}^{-3}}
\newcommand{\pc}{\rm{pc}}
\newcommand{\kpc}{\rm{kpc}}
\newcommand{\mearth}{M_\oplus}
\newcommand{\mjup}{M_J}
\newcommand{\rearth}{R_\oplus}
\newcommand{\rjup}{R_J}
\newcommand{\au}{\rm{AU}}
\newcommand{\yr}{\rm{yr}}

\newcommand\aj{{AJ}}
\newcommand\araa{{ARA\&A}}
\newcommand\apj{{ApJ}}
\newcommand\apjl{{ApJ}}
\newcommand\apjs{{ApJS}}
\newcommand\aap{{A\&A}}
\newcommand\aaps{{A\&AS}}
\newcommand\mnras{{MNRAS}}
\newcommand\nat{{Nature}}
\newcommand\jgr{{J.~Geophys.~Res.}}

\newcommand\icarus{{Icarus}}

\title[Planets in clusters]
  {Planets in Open Clusters Detectable by Kepler}
\author[S. Chatterjee et al.]
  {Sourav~Chatterjee,$^1$\thanks{email: s.chatterjee@astro.ufl.edu}
  Eric~B.~Ford,$^1$ Aaron~M.~Geller,$^2$ Frederic~A.~Rasio$^{2}$\\
  $^1$Department of Astronomy, University of Florida, Gainesville, FL 32611\\
  $^2$Center for Interdisciplinary Exploration and Research in Astrophysics (CIERA) and \\ 
  Dept. of Physics \& Astronomy, Northwestern University, 2145 Sheridan Rd, Evanston, IL 60208, USA.\\
  }

\pagerange{\pageref{firstpage}--\pageref{lastpage}} \pubyear{2002}

\def\LaTeX{L\kern-.36em\raise.3ex\hbox{a}\kern-.15em
    T\kern-.1667em\lower.7ex\hbox{E}\kern-.125emX}

\begin{document}

\label{firstpage}

\maketitle

\begin{abstract}
 
\end{abstract}
While hundreds of planets have been discovered around field stars, only a few 
are known in star clusters.  To explain the lack of short-period giant planets 
in globular clusters (GC), such as 47~Tucane and $\omega$~Centauri, it 
has been suggested that their low metallicities may have prevented planet formation.  
Alternatively, the high rates of close stellar encounters in these 
clusters may have influenced the formation and subsequent evolution of planetary systems.  
How common are planets in clusters around normal main-sequence stars?  
Here we consider whether this question can be addressed using data from the 
Kepler mission.  The Kepler field of view contains 4 low-density (relative to GCs) open clusters where 
the metallicities are about solar (or even higher) and stellar encounters 
are much less frequent than in typical GCs.  We provide detailed $N$-body models 
and show that most planets in Kepler-detectable orbits are not significantly perturbed 
by stellar encounters in these open clusters.  We focus on the most 
massive cluster, NGC~6791, which has super-solar metallicity, and find 
that if planets formed in this cluster at the same frequency as observed in the field, 
Kepler could detect 1 -- 20 transiting planets depending on the planet-size distribution and the duration 
of data collection.  
Due to the large distance to NGC~6791 Kepler will have to search relatively faint ($K_p<20$) stars 
for the full extended mission to achieve such a yield.  

\section{Introduction}
\label{sec:intro}
Stars are thought to always form in clusters or groups 
\citep[e.g., ][]{2000AJ....120.3139C,2003ARA&A..41...57L,2003AJ....126.1916P}. Most of these
clusters dissolve rather quickly under Galactic tides and via encounters with giant molecular clouds, 
forming the apparently isolated
stars in the field.  Multiple independent constraints indicate that our now-isolated star, the Sun, was born in a
moderately-sized star cluster with $N = 10^3$ -- $10^4$ stars
\citep{2010ARA&A..48...47A}.  However, this understanding leads to an apparent
puzzle.  Planets within a few AU from the host stars
are observed to be common around normal main sequence (MS) stars in the field
\citep[e.g.,][]{2006ApJ...646..505B,2012arXiv1202.5852B}. In contrast, few planets 
have been discovered around MS stars in star clusters. 
The only planets confirmed in star clusters to date are the 
Jovian-mass planet in a wide circumbinary orbit around a white-dwarf and a millisecond pulsar in 
M4 \citep[e.g., ][]{2003Sci...301..193S,2006ApJ...640.1086F}, the giant planet of minimum mass $7.6\,\mjup$ around 
$\epsilon$~Tauri, a giant star in the open cluster Hyades \citep{2007ApJ...661..527S}, and most recently 
the two hot jupiter planets around F-G stars Pr0201 and Pr0211 in Praesepe \citep{2012arXiv1207.0818Q}.
Although a number of transiting planet candidates have been 
reported \citep[e.g.,][]{2006AJ....131.1090M,2007A&A...472..657L,2008MNRAS.387..349M,2008AJ....135..907P,2003A&A...410..323B} 
around clusters stars, the planets recently discovered in Praesepe are the only planets discovered around 
MS stars in a star cluster.
Do planets not form around cluster stars as often as around field stars?  If they do form, does
some physical process such as stellar encounters destroy planets with detectable orbits?  Or have
searches simply not been sensitive enough? 

Massive ($\geq 10^4\,\msun$) and dense clusters like typical
Galactic GCs have low metallicities \citep{1996AJ....112.1487H}.
Radial velocity (RV) surveys of field stars show that higher metallicity stars are more likely to host short-period giant
planets \citep{2005ApJ...622.1102F}.  Thus, it has been suggested
that the low metallicities of Galactic GCs may inhibit
planet formation around stars in these clusters.  Nevertheless, recent theoretical
studies suggest that planets may be able to form even
at such low metallicities, especially in short-period orbits
\citep[$Z \geq 0.1 Z_\odot$;][]{2008IAUS..249..325H,2012arXiv1203.4817J}.
Alternatively, the high interaction rates in the relatively
crowded regions of star clusters may have led to the destruction of detectable planetary orbits (or
close encounters may have inhibited their formation via, e.g., disk truncation).  In contrast to the 
GCs, open clusters (typically $N\sim10^2$--$10^3$) have much lower central stellar densities. 
Analytical estimates indicate that in these 
lower mass clusters stellar encounters should not affect planet formation and 
their subsequent survival, especially for planets in orbits detectable by transit 
searches \citep[e.g., ][]{2001Icar..153..416K,2006ApJ...641..504A}. Of course the interaction rate 
depends on the planet's semimajor axis, and a true semimajor-axis-dependent encounter frequency 
can only be obtained via direct $N$-body integration of a star cluster with stars hosting planets 
with a large range of initial semimajor axes. 

Finding a planet around a star in a low-mass cluster is challenging.  These clusters dissolve
at a young age under the Galactic tidal forces \citep{2003MNRAS.340..227B}.  Therefore, the
observation must be made at a relatively early age before the cluster dissolves. High precision 
Doppler or transit observations are not easy for young stars due to high level of star spots and stellar activity. 
In addition, ongoing star formation, gas expulsion, and presence of high-mass ($M\geq10\,\msun$) stars in young clusters 
make these observations hard. 
Hence, an ideal star cluster for a transit search for planets should have the following properties:
(1) high enough metallicity for planets to form; (2) a sufficiently
massive cluster so that it does not dissolve too quickly; (3) a
sufficiently old cluster so that both star and planet formation have been completed. 

Searching for planets in star clusters using
transit surveys has been suggested to take advantage of the high surface density
of target stars \citep{1996JGR...10114853J}. Moreover, finding the frequency of planets 
around cluster stars provides us with useful constraints on how often planets (within a range 
of semimajor axes depending on the detection limit) form, migrate, and survive around a 
coeval aggregate of stars. 
Transit surveys have been used to search for hot Jupiters in 47~Tuc, yielding
a significant null result \citep{2000ApJ...545L..47G,2005ApJ...620.1043W}.
A similar null result was found for $\omega$~Cen
\citep{2008ApJ...674.1117W}.
A number of transit searches in various open clusters also revealed no convincing
planet candidates around normal MS stars
\citep[e.g., ][]{2006AJ....132..210B,2009ApJ...695..336H,2006AJ....132..210B,2005AJ....129.2856M,2006AJ....131.1090M,2008MNRAS.387..349M,2005MNRAS.359.1096B,2006MNRAS.367.1677B,2007A&A...470.1137M,2009A&A...505.1129M,2006AJ....132.2309R}.
Nevertheless, these null results from ground based surveys do not yet
constrain the planet occurrence rate in
open clusters to be lower than that expected from observations in the field \citep{2011ApJ...729...63V}. 
Hence, it would be very interesting to identify a promising candidate cluster where planets {\em are} expected 
to be detected. The discovery of planets in such a cluster would provide us answers to the apparent 
puzzle created by the presumed cluster origin of all stars, high frequency of short period planets around field stars, 
and the apparent lack of short period planets around cluster stars. Alternatively, a null result from that 
cluster would potentially put stronger constraints on planet formation efficiency in clusters 
than are currently present. NASA's Kepler mission provides an unprecedented opportunity to investigate 
this problem due to its nearly uninterrupted photometric measurements with high cadence and precision 
over many years.     

Four open clusters lie within the Kepler field of view (FOV), namely NGC~6866, NGC~6811, NGC~6819, and 
NGC~6791, in increasing order of their ages \citep{2011ApJ...733L...9M}. All 4 clusters have solar or 
higher metallicities 
\citep[e.g.,][]{2011ApJ...733L...1P,2009AJ....138..159H,2012arXiv1205.2760G,2012arXiv1205.4023C}. 
Among these 4 clusters, NGC~6791 is the most massive ($M_{\rm{tot}}\approx5\times10^3\,\msun$, \citealt{2011ApJ...733L...1P}) 
and provides the largest number of targets to 
search for planets transiting normal MS stars.
NGC~6791 is old ($8\pm1\,\gyr$; \citealt{2008A&A...492..171G,2011ApJ...733L...1P,2011ApJ...733L...9M,2012A&A...543A.106B}), has 
super-solar metallicity ([Fe/H] = $+0.30$; \citealt{2009AJ....137.4949B}) and its central 
density is relatively low ($\rho_c \approx 60\,\dens$; \citealt{2011ApJ...733L...1P}), much lower than in any GC. 
The high metallicity indicates that planet formation in NGC~6791 should not be inhibited 
due to lack of metals, and the low stellar density relative to typical GCs indicates that 
stellar encounters are not as common as in the GCs. 

In this study we use detailed $N$-body simulations with planet-harboring stars and stellar 
binaries incorporating stellar evolution, two-body relaxation, Galactic tidal stripping, dynamical 
interactions between stars, binaries, and planetary systems, and physical collisions. 
Our goal is to study the semimajor axis-dependent effects of stellar encounters on planetary orbits 
in clusters similar to NGC~6791. Moreover, using initial distributions of planet properties based on 
the Kepler planet candidate list we estimate the number and properties of planets that Kepler 
can detect in NGC~6791 if the planet occurrence rate in this cluster is the same as that in the 
field.  

Our numerical methods are summarized in Section\ \ref{sec:models}.  We present our model 
of NGC~6791 in Section\ \ref{sec:ngc6791}.  
We then explore various properties indicating the effects of stellar encounters
on the planetary orbits in our cluster models (Section\ \ref{sec:planets}) with particular focus on
the model that best matches the observed properties of NGC~6791.
In Section\ \ref{sec:detect} we investigate whether planets in NGC~6791 can be detected by Kepler 
and discuss the expected properties of these detectable planets.
We conclude and discuss the implications of our results in Section\ \ref{sec:conclusion}.

\section{Numerical Methods}
\label{sec:models}
In this section we describe the code we use, how we assign cluster properties (e.g., mass, concentration, 
stellar binary fraction, and fraction of planet-harboring stars; Section\ \ref{sec:modeling_ngc6791}), 
planetary masses and their orbital properties (Section\ \ref{sec:modeling_planet}).  
We describe how we estimate the expected number of Kepler-detectable planets in Section\ \ref{sec:modeling_signal}.  
\subsection{Modeling of NGC~6791}
\label{sec:modeling_ngc6791}
We model star clusters with planet hosting stars using a H\'enon-type Monte Carlo 
code CMC (for cluster Monte Carlo) developed and rigorously tested over the past decade 
\citep{2000ApJ...540..969J,2001ApJ...550..691J,2003ApJ...593..772F,2007ApJ...658.1047F,2010ApJ...719..915C,2012ApJ...750...31U}.  Using this code 
we can efficiently model realistic clusters with a large number of star-systems ($N$) 
and high binary fraction ($f_b$) 
including two-body relaxation, binary and single stellar evolution, strong encounters 
(including physical collisions and binary mediated scattering interactions), and Galactic 
tidal stripping at a relatively low computational cost.  
These models can be directly compared with observed star clusters.      

The collective effect of all of the above physical processes over the age of the cluster 
determines the properties of an observed cluster.  
Although the qualitative effects of each physical process is known, the quantitative 
extent of these effects are hard to estimate without doing full simulations.  
Hence, the initial conditions necessary to create an observed cluster cannot be 
easily inverted.  We perform simulations for initial conditions spanning a large grid of multi-dimensional 
parameter space.  
We focus on modeling NGC~6791 and summarize a collection of properties for all 
other models as representative of rich open clusters.  
We restrict the huge parameter space to make calculations tractable.  
For example, NGC~6791 has a Galactocentric distance between $5$ and $10\,\kpc$ 
\citep[e.g., ][]{2011ApJ...733L...1P}.  It is also expected that the cluster loses about 
$50$ -- $70\%$ of its initial mass via Galactic tidal stripping, dynamical ejections, and 
stellar evolution mass loss through winds and compact object formation 
\citep[e.g., ][]{2003MNRAS.340..227B,2010ApJ...719..915C}.  Considering these 
{\em a priori} constraints we compute a large grid of $\sim 200$ simulations varying the initial parameters 
including the cluster mass ($M_{\rm{cl}}$), King concentration parameter 
($W_0$), Galactocentric distance ($r_{\rm{GC}}$), 
and virial radius ($r_v$).  We compare the simulated cluster properties between 
$7$--$9\,\gyr$ with the observed properties including the surface number density profile 
of NGC~6791 to find acceptable matches.  

All our simulated clusters have an initial $N$ between $10^4$ -- $10^5$ and an initial 
$r_v$ between $3$--$8\,\pc$.  The positions and velocities are 
drawn according to a King model \citep{1966AJ.....71...64K,2008gady.book.....B} 
with $W_0$ between $3$ -- $6$.  The initial stellar binary 
fraction ($f_b$) is chosen to be between $0.1$ and $0.5$.  
We draw the masses of the stars (or primary stars in binaries) 
from the stellar mass function (MF) presented in \citet[][Equations\ $1$ and $2$]{2001MNRAS.322..231K} 
in the stellar mass range $0.1-100\,\msun$.  
The mass of each secondary in a stellar binary 
is drawn from a uniform distribution of mass ratios in the range 
$0.1\,\msun$ --  the mass of the primary.  
The initial eccentricities ($e_i$) are chosen to be thermal \citep{2003gmbp.book.....H}.  
The initial semimajor axis ($a_i$) distribution 
is flat in logarithmic intervals between $5$ times physical contact to the local 
hard-soft boundary of the cluster.  
The local hard-soft boundary is a measure to determine the widest orbit of a stellar binary that 
will not be disrupted via stellar encounters in its cluster environment and depends on the velocity 
dispersion of the stars in 
the region of the cluster and the binding energy of the binary.  The ``hard" binaries are sufficiently bound 
to each other and statistically becomes more bound via super-elastic stellar encounters.  On the other 
hand ``soft" binaries are those that are not sufficiently bound and becomes less bound via stellar encounters 
and eventually gets disrupted via one (or multiple) stellar encounters \citep{2003gmbp.book.....H}.  
Although all binaries are initially hard, they may not remain so throughout the evolution of the cluster.  
Due to two-body relaxation the velocity dispersion near the center of the cluster increases as the core 
contracts.  Moreover, binaries sink to the center due to mass segregation where the velocity dispersion is 
higher compared to that at the initial position of a binary.  We include this effect in our simulations and 
these soft binaries are allowed in the cluster until they disrupt via binary-single or binary-binary interactions.  
The orbital phase angles and orientations are 
chosen uniformly in the full range.  

\subsection{Planet properties}
\label{sec:modeling_planet}
In addition to stellar binaries we also include stars with planetary companions determined 
by a planet fraction $f_p$ defined as the ratio of the number of planet host stars ($N_p$) to 
the number of all star systems (singles or binaries; $N$) in the cluster.  Each primary 
can have only one companion in our simulations, either a stellar mass companion (we call 
those stellar binaries) or a planetary companion (we call those planetary binaries).  Strong 
interactions involving both stellar binaries and planetary binaries in the evolving cluster 
potential are followed.  In any simulation the total fraction of ``binaries" (stellar and planetary) 
is $f_b + f_p$.  We do not include multiple planet systems or circumbinary (i.e., orbiting a binary star) 
planets  due to a limitation in the current 
version of the code that does not allow us to treat hierarchical systems above a binary.  

Planet masses ($M_p$) are assigned according to a power-law $df/d{\rm{log}}M_p = M_p^{-0.48}$ 
\citep{2010Sci...330..653H} between $M_p = 1\,\mearth$ -- $5\,\mjup$ where, 
$\mjup$ is the mass of Jupiter.  The planetary radii ($R_p$) are assigned a mass-dependent 
value according to $R_p =$ the smaller of $\rearth \left(\frac{M_p}{\mearth}\right)^{\frac{1}{2.06}}$ and 
$\rjup$, where $\rjup$ is the radius of Jupiter \citep{2011arXiv1102.0543L}.  
Note that although planets with $R_p > \rjup$ are 
observed, we employ the upper limit in $R_p$ to remain conservative in our estimated 
number of Kepler-detectable planets (described later in Section\ \ref{sec:modeling_signal}).  

We assign the planetary orbits in two different ways, henceforth {\tt Set1} and {\tt Set2}.  In 
{\tt Set1} our focus is to find the effects of stellar encounters on planetary orbits as a function of 
the planetary semimajor axes ($a_p$) in a cluster environment.  The encounter rate of a binary in a star cluster is directly 
dependent on its semimajor axis $a$, and is proportional to $a^2$.  Hence, to estimate the $a_p$-dependent 
effect of stellar encounters on planetary orbits we need to sample $a_p$ over a large range.    
For {\tt Set1} the initial $a_p$-distribution is flat in logarithmic intervals between $10^{-2}$ and 
$10^2\,\au$.  For {\tt Set1} we use a fixed value of $f_p = 0.33$, a somewhat arbitrary choice, but 
not too far from the overall fraction of Kepler planet candidates observed in the field.    

{\tt Set2} consists of models with initial conditions resulting in a close match with NGC~6791 
from the large grid of simulations in {\tt Set1}.  Here we focus on estimating the expected 
number of planets detectable by Kepler in NGC~6791 assuming that the initial planet frequency in 
the cluster is the same as observed in the field.  
In {\tt Set2} the orbital period distribution for planets is guided by the observed period distribution 
of the Kepler planet candidates \citep[][]{2012arXiv1202.5852B}.  Each Kepler planet candidate 
is weighted by $a/R_*$, where $R_*$ is the radius of the host star, to account for the geometric 
transit probability.  A lognormal distribution for orbital period is obtained by fitting the weighted 
Kepler planet candidate period distribution.  
We draw initial planet orbital periods from this lognormal distribution between $3$ and 
$365$ days for {\tt Set2}.  We use $f_p = 45\%$ for {\tt Set2}.  This 
value of $f_p$ is obtained requiring that the rate of transiting planets 
with $R_p > 2.5\,\rearth$ and period between $3$ and $120$ days is equal 
to the rate of Kepler planet candidates observed in the field.  The Kepler planet 
candidate list should suffer from only minor incompleteness for planets in this range around 
dwarf host stars with $K_p<14$ \citep{2012arXiv1202.5852B}.         

In all cases the initial planetary orbits are circular.  Since we do not include multiple planets, 
there is no excitation of orbits via planet-planet scattering \citep[e.g., ][]{2008ApJ...686..580C} 
in our simulations.  Thus, any final non-zero $e$ is a result of stellar encounters in the cluster.  
In (Section\ \ref{sec:encounter_orbit}) we argue that indirect planet-planet instabilities, i.e., planet-planet 
scattering triggered by stellar fly-bys, should be a small 
percentage effect in clusters like NGC~6791.  The orbital phase and orientation angles are drawn uniformly in the 
full range in both sets.  
Planetary companions are assigned only around host stars with $M_\star \leq 2\,\msun$ 
in both sets since transiting planet searches focus on these systems.  

\subsection{Detectability of Planet Transits by Kepler}
\label{sec:modeling_signal}
In our simulations we track single and binary stellar evolution in tandem with the dynamics 
\citep[e.g.,][]{2010ApJ...719..915C}.  Thus, at any given time in the evolution, we 
can extract the stellar radii ($R_\star$) and bolometric luminosities ($L_\star$) of planet host stars 
and the orbital properties of the planets.  The bolometric luminosities are converted to standard 
colors using the standard filter band passes (Johnson $B$, $V$) and synthetic stellar spectra obtained 
from stellar atmospheric models dependent on the stellar metallicities, and surface gravity 
\citep{1997yCat..41250229L,1997A&AS..125..229L}.  The $B$, $V$ band magnitudes are then converted 
to SDSS $g$, and $r$ magnitudes using the transformation equations 
from \citet{2002AJ....123.2121S}.  The Kepler magnitude $K_p$ is then calculated using the 
standard Kepler conversions obtained from the Guest Observer (GO) website of Kepler 
(http://keplergo.arc.nasa.gov/CalibrationZeropoint.shtml).  Note that the conversion equations 
in the Kepler GO website is slightly different from those given in \citet{2011AJ....142..112B}.  We find that 
the $K_p$ values calculated using the conversion equations in the Kepler GO website match 
better with the actual $K_p$ values in the Kepler target list (lower rms difference) compared 
to the $K_p$ values calculated using the equations given in \citet{2011AJ....142..112B}.       

Whether or not a planet would be detected by Kepler depends on the ratio of the 
transit signal strength and the noise metric from all contributing factors (including photon noise and 
intrinsic stellar 
variability) for Kepler in a specified temporal length for the host.  For a noise metric we use 
the combined differential photometric precision (CDPP) measured over 6.5 hours.   
We assign the CDPP values in ppm for the planet host stars in our simulations using a polynomial fit: 
\begin{equation}
\label{eq:cdpp}
\log CDPP(K_p) = 10.488 - 2.9878 Kp + 0.34182 Kp^2 - 0.017045 Kp^3 + 0.00033146 Kp^4
\end{equation}
to the CDPP values measured by Kepler and given in \citet[][their Figure\ 4]{2011ApJS..197....6G}.  
Note that in general extrapolating the CDPP values to large $K_p$ values ($K_p = 20$ needed in 
our study) using CDPP data in the range $K_p < 16$ \citep[given in ][]{2011ApJS..197....6G} can be 
dangerous.  However, an investigation by Gilliland (private communication with the referee Gilliland) 
unveils that the CDPP values predicted using this extrapolation equation indeed provides CDPP 
values in acceptable agreement with real data even for $K_p = 20$.  The values predicted by the 
extrapolation equation provides a more conservative estimate compared to the real data (private 
communication with the referee, Gilliland).  
The signal to noise ratio (SNR) of a transiting planet is calculated for each planet using 
\begin{equation}
\label{eq:snr}
SNR = \frac{(R_p/R_\star)^2}{CDPP} \left( \frac{n_{\rm{tr}} t_{\rm{dur}}}{6.5\,\rm{hr}} \right)^{0.5}
\end{equation}       
\citep{2012ApJS..201...15H}, where $n_{\rm{tr}}$ is the number of transit events within a given 
length of data collection by Kepler (e.g., $1$, $3.5$, and $8\,\rm{yr}$ are considered), and 
$t_{\rm{dur}}$ is the duration of transit, given by  
\begin{equation}
\label{eq:t_dur}
t_{\rm{dur}} = 2 R_\star \left( \frac{a}{GM_\star} \right)^{0.5}
\end{equation}
assuming a circular orbit and central transit.  
If the $SNR>7$, we consider the planetary orbit to be ``detectable" by Kepler.  Each 
detectable planet is then weighted by the geometrical transit probability 
$P_{\rm{tr}} = R_\star / a$ to estimate the actual number of transiting planets detected 
by a single observer.  

The predicted number of detectable transiting planets can vary due to 
statistical fluctuations in their periods, sizes, and which planets were assigned to which stars in our simulations.  
Therefore, we vary the seed of the random number generator and generate $4$ 
realizations of the initial conditions for each set of cluster input parameters 
that gives us a good match for NGC~6791 to estimate the size of the 
statistical fluctuations.  Note that the estimated CDPP values for a given $K_p$ is obtained for 
stars on the MS \citep{2011ApJS..197....6G}.  If the CDPP values are significantly different around 
giant stars, then the noise metric estimated using Equation\ \ref{eq:cdpp} may be inadequate.  
Nevertheless, we will show that the majority of Kepler detectable transiting planets are around bright 
MS stars (and some subgiant stars), hence, this caveat is not expected to be serious for the purpose of this study.      

\section{Simulated model of NGC~6791}
\label{sec:ngc6791}
%
\begin{center}
\begin{longtable}{lccccc}
\caption{List of properties for our simulated model of NGC~6791.  The initial 
parameters are given as well as the properties of the model cluster at $8\,\gyr$.  Two columns for the 
final properties are based on including all stars in the model, and a subset of stars satisfying $g<22$.  
The observed properties are taken from \citet{2011ApJ...733L...1P} for comparison.  } \label{tab:NGC6791} \\   
\hline \hline \\
\multicolumn{1}{l}{\textbf{Cluster Property}} & 
\multicolumn{1}{c}{\textbf{Initial Value}} & 
\multicolumn{2}{c}{\textbf{Final Value}} & 
\multicolumn{1}{c}{\textbf{Observed Value}} \\ \cline{2-5}
\multicolumn{1}{c}{} &
\multicolumn{1}{c}{\textbf{}} &
\multicolumn{1}{c}{\textbf{All Star}} &
\multicolumn{1}{c}{\textbf{$g<22$}} &
\multicolumn{1}{c}{\textbf{$g<22$}} \\
\hline \hline\\
\endhead
%

\endfoot

$N$ & $8\times10^4$ & $1.8\times10^4$ & $3.8\times10^2$ & - \\
Total Cluster Mass ($\msun$) & $6\times10^4$ & $10^4$ & $4.5\times10^3$ & $5\times10^3$ \\
Concentration parameter ($\log\frac{r_t}{r_c}$) & $1.1$ & $0.91$ & - & $0.9$ \\
Virial Radius & $8\,\pc$ & - & - & - \\
Galactocentric Distance ($\kpc$) & $8.5$ & $8.5$ &  & $5$ -- $10$\\
Stellar binary fraction ($f_b$) & $0.30$ & $0.35$ &  & - \\
Fraction of planet-harboring stars ($f_p$) & $0.33$ & $0.27$ &  & -\\
\hline \hline
\end{longtable}
\end{center}
%
%
\begin{figure}
\begin{center}
\includegraphics[width=0.9\textwidth]{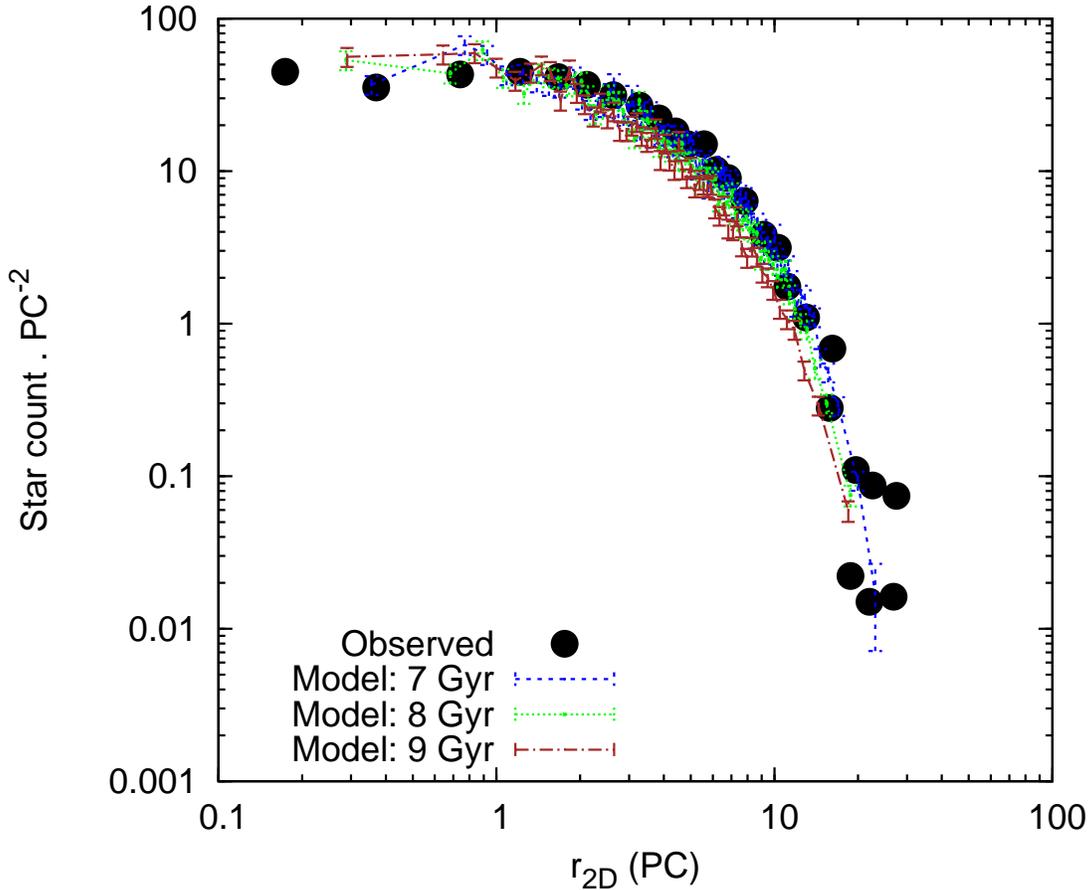}
\caption[Comparison:Surface density profile]{Surface density profile for the observed cluster 
(black dots) and the best-match simulated cluster at three times between 
$t_{\rm{cl}} = 8 \pm 1\,\gyr$ (blue, green, and brown in increasing order of $t_{\rm{cl}}$) during its evolution.  
For the simulated cluster only stars 
with $g < 22$ are considered to be 
compatible with the observed data.  The observed data is obtained using the Dexter data 
extraction applet \citep{2001ASPC..238..321D} from \citet{2011ApJ...733L...1P}.  One of the key 
questions for this study is how much planetary orbits are perturbed in a cluster environment.  This is 
directly dependent on the density profile of the cluster.  The close similarity in the density 
profiles for the observed NGC~6791 and our theoretical models indicate that the dynamical 
interaction rate in our model should be very similar to that in NGC~6791.    
  }
\label{plot:sbp}
\end{center}
\end{figure} 

To compare our models from {\tt Set1} with the observed NGC~6791 we need to ``observe" 
our cluster model.  We compare the surface number density profiles from our models, a basic observable for 
clusters, with that of NGC~6791 at $t_{\rm{cl}} = 8 \pm 1\,\gyr$ \citep{2011ApJ...733L...1P}.  
In order to compare 
with the observed cluster we convert the bolometric luminosities obtained directly from CMC 
to standard colors using the standard filter band passes 
and synthetic stellar spectra obtained from stellar atmospheric models dependent on the stellar metallicities, 
and surface gravity \citep{1997yCat..41250229L,1997A&AS..125..229L}.  We only include objects 
brighter than $g < 22$ to create the 
surface density profiles.  A good match in the surface brightness profile indicates 
that the total luminosity (a proxy for mass) of the cluster as well as its radial distribution is well modeled. 
Indeed, in observed clusters the surface density profile is often used to estimate many 
global structural properties including central density ($\rho_c$), 
core radius ($r_c$), and half-light radius ($r_{hl}$).  

Note that for this study we do not compare various stellar populations.  Clusters are birth places for 
a number of exotic stellar populations including blue straggler stars (which should be created mostly 
via mass-transfer in a stellar binary given the low central density of NGC~6791), and low-mass X-ray binaries 
\citep{2006ApJ...646L.143P}.  These populations depend closely on the details of the initial stellar MF, 
stellar binary properties, the stellar density profile, and the dynamical evolution of the cluster.  While 
numerical modeling of star clusters focusing on reproduction of these individual exotic stellar 
populations is a very interesting and active area of research, it is beyond the scope of this study.  
For our purposes, the most important aspect in a cluster environment is its density distribution since 
the stellar density distribution directly affects the interaction rates and in turn affects how the 
properties of planetary orbits are expected to change in the cluster.  

\begin{figure}
\begin{center}
\includegraphics[width=0.9\textwidth]{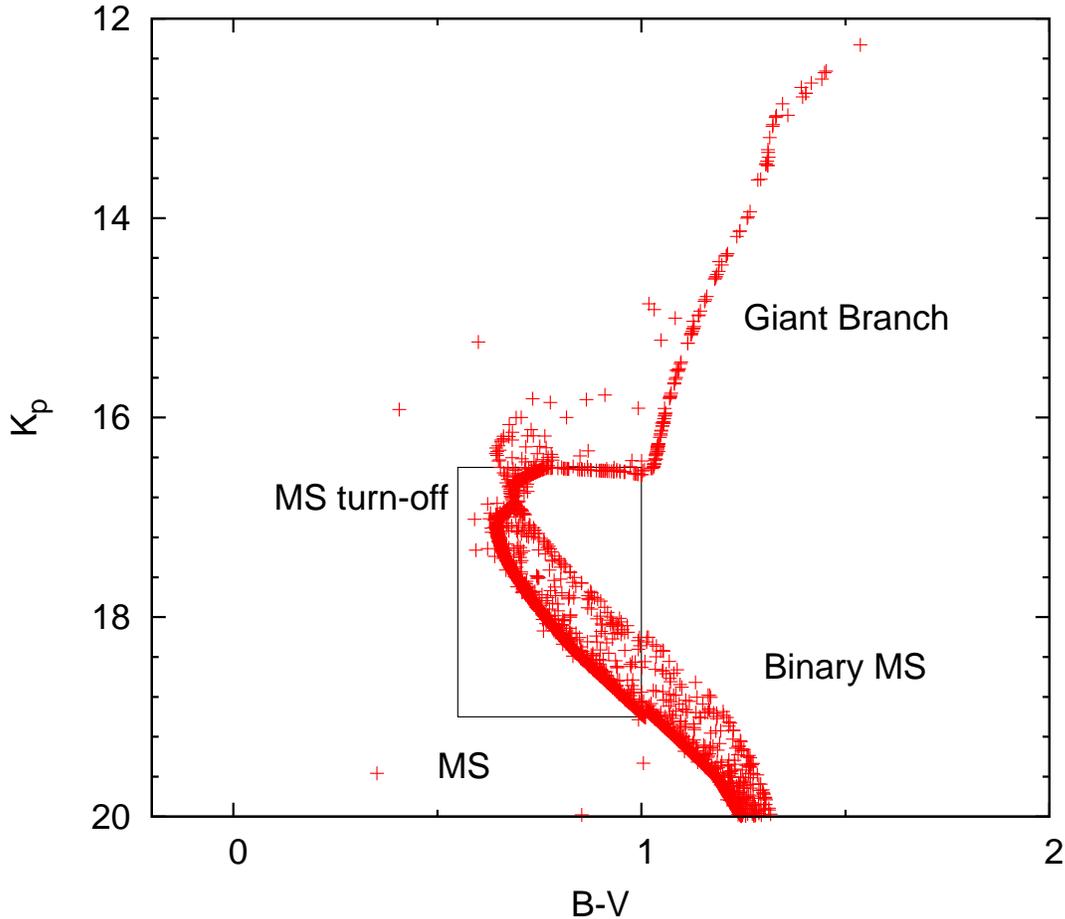}
\caption[HRD:$K_p$ vs B-V]{A synthetic CMD for the best-match cluster model for NGC~6791.  
The single and binary MSs are clearly seen as well as the red giant branch.  
We find that the best chance for detecting planets is around stars residing near the high 
end of the MS (box).  The giants are brighter, but they are also much bigger (the radius 
can be $~10^2$ times compared to the radius of the star during MS) resulting in a drop in 
the signal to noise ratio for Kepler.  The number of blue stragglers \citep{1953AJ.....58...61S} 
in our model is lower compared to what is observed in NGC~6791.  The blue stragglers in NGC~6791 
are expected to be created via mass transfer in binaries due to the low central density in 
the cluster and should not be indicative of the dynamical interaction rate which is dependent on 
the density profile of the cluster (Figure\ \ref{plot:sbp}).  
  }
\label{plot:hrd}
\end{center}
\end{figure} 
The initial conditions that produce a good match 
with NGC~6791 are presented in Table\ \ref{tab:NGC6791}.  Table\ \ref{tab:NGC6791} 
also lists the final properties of the model cluster at age $t = 8\,\gyr$ and some observed 
properties of NGC~6791.
Figure\ \ref{plot:sbp} shows the comparison between the observed surface density profile 
of NGC~6791 and that for our simulated best-match model at three different snapshots 
between $t_{\rm{cl}} = 8 \pm 1\,\gyr$.  

By tracking both the dynamical evolution of the cluster objects 
in the overall cluster potential and the stellar and binary properties as they evolve we can 
create realistic synthetic color-magnitude diagrams (CMD) for our simulated models.  
Figure\ \ref{plot:hrd} shows an example of synthetic CMD created 
for the cluster model showing best match for the observed NGC~6791 at $t_{\rm{cl}} = 8\,\gyr$.  
The single 
and binary MSs, the MS turn-off, and the Giant branch are clearly visible.  
NGC~6791 is old \citep[about $8\,\gyr$;][]{2008A&A...492..171G} 
and it is relatively far from us \citep[$4\,\kpc$;][]{2011ApJ...733L...1P}.  
Hence, the MS stars in this cluster are not very bright.  
Note that the single MS turn-off is at $\approx 17\,K_p$ and the binary MS turn-off is near $16\,K_p$.        

We summarize some key properties of all our simulated clusters in the wide multidimensional 
grid from our search in Table\ \ref{tab:list} available in its entirety in the electronic version.  
The best-match model for NGC~6791 is listed as {\bf run167} in the Table\ \ref{tab:list}.  
In the rest of this study we will focus on the results using these cluster parameters.  
\begin{center}
\begin{longtable}{ccccccccccccccccccccc}
\caption{List of properties for all our simulated cluster models at different cluster ages ($t_{cl}$).  This table 
is presented in its entirety in the online journal.  A portion is presented here for guidance to its form 
and content.  The 
initial parameters for each model including the King concentration parameter $W_0$, virial radius $r_v$, 
Galactocentric distance $r_{GC}$ are given below each run name.  Various properties are listed at a few 
$t_{cl}$ values.  The total cluster mass $M$, total cluster object number $N$, central density $\rho_c$, 
stellar binary fraction $f_b$, fraction of planet host stars $f_p$ are listed.  The ratio of the number of 
free-floating planets bound to the cluster ($N_{p, ff, bound}$) to the number of planets bound to their host stars 
in long orbits $N_p(a>10\,\au$) is denoted as $f_{p, ff, bound}$.  The ratio between $N_p$ and the number of planets 
escaped from the cluster potential and still bound to their hosts $N_{p, esc}$ is denoted by $f_{p, esc}$.  
The ratio between $N_{p, esc}$ and the number of free-floating planets escaped from the cluster potential 
($N_{p, ff, esc}$) is denoted by $f_{p, ff, esc}$.  Cluster models that dissolve due to Galactic tides 
within a Gyr because of their low mass and compactness are excluded from the list.  } \label{tab:list} \\   
\hline \hline
\multicolumn{1}{c}{\textbf{Name}} & 
\multicolumn{1}{c}{\textbf{$t_{cl} (\gyr)$}} & 
\multicolumn{1}{c}{\textbf{$M$ ($10^3\,\msun$)}} & 
\multicolumn{1}{c}{\textbf{$N/10^3$}} & 
\multicolumn{1}{c}{\textbf{$\rho_c$ ($\msun \rm{pc^{-3}}$)}} &
\multicolumn{1}{c}{\textbf{$f_b$}} &
\multicolumn{1}{c}{\textbf{$f_p$}} &
\multicolumn{1}{c}{\textbf{$f_{p, ff, bound}$}} &
\multicolumn{1}{c}{\textbf{$f_{p, esc}$}} & 
\multicolumn{1}{c}{\textbf{$f_{p, ff, esc}$}} \\
\hline \hline
\endfirsthead

\hline \hline
\multicolumn{1}{c}{\textbf{Name}} & 
\multicolumn{1}{c}{\textbf{$t_{cl}\,\gyr$}} & 
\multicolumn{1}{c}{\textbf{$M$ ($10^3\,\msun$)}} & 
\multicolumn{1}{c}{\textbf{$N/10^3$}} & 
\multicolumn{1}{c}{\textbf{$\rho_c$ ($\msun \rm{pc^{-3}}$)}} &
\multicolumn{1}{c}{\textbf{$f_b$}} &
\multicolumn{1}{c}{\textbf{$f_p$}} &
\multicolumn{1}{c}{\textbf{$f_{p, ff, bound}$}} &
\multicolumn{1}{c}{\textbf{$f_{p, esc}$}} & 
\multicolumn{1}{c}{\textbf{$f_{p, ff, esc}$}} \\
\hline \hline
\endhead

\multicolumn{21}{l}{{Continued on Next Page\ldots}} \\
\endfoot
\\[-1.8ex] \hline \hline 
\endlastfoot
{\bf run1} & 0.00 & 13.4 & 20.0 & 392.7 & 0.10 & 0.33 & 0.0006 & 0 & - \\
$W_0=3$ & 0.27 & 9.1 & 19.2 & 13.7 & 0.10 & 0.33 & 0.001 & 0.03 & 0 \\
$r_v=5.0\,\pc$ & 0.55 & 8.3 & 18.0 & 11.4 & 0.10 & 0.33 & 0.002 & 0.1 & 0.002 \\
$r_{GC}=7\,\kpc$ & 0.82 & 7.6 & 16.8 & 10.7 & 0.10 & 0.33 & 0.002 & 0.2 & 0.006 \\
  & 1.09 & 6.9 & 15.2 & 10.1 & 0.10 & 0.33 & 0.004 & 0.3 & 0.005 \\
  & 1.36 & 6.3 & 13.6 & 9.8 & 0.10 & 0.33 & 0.004 & 0.5 & 0.005 \\
  & 1.64 & 5.5 & 11.6 & 7.4 & 0.11 & 0.32 & 0.002 & 0.7 & 0.005 \\
  & 1.90 & 4.7 & 9.6 & 8.0 & 0.11 & 0.32 & 0.005 & 1 & 0.005 \\
  & 2.17 & 3.8 & 7.3 & 9.0 & 0.12 & 0.32 & 0.005 & 2 & 0.004 \\
  & 2.45 & 2.9 & 5.3 & 9.0 & 0.12 & 0.31 & 0.009 & 3 & 0.004 \\
  & 2.72 & 2.2 & 3.6 & 9.5 & 0.13 & 0.31 & 0.01 & 5 & 0.004 \\
\hline
{\bf run2} & 0.00 & 27.1 & 40.0 & 657.9 & 0.10 & 0.33 & 0 & 0 & - \\
$W_0=3$ & 0.34 & 18.4 & 39.3 & 27.4 & 0.10 & 0.33 & 0.005 & 0.01 & 0.07 \\
$r_v=5.0$ pc & 0.68 & 17.4 & 38.6 & 21.0 & 0.10 & 0.33 & 0.009 & 0.03 & 0.05 \\
$r_{GC}=7$ kpc & 1.03 & 16.6 & 37.6 & 18.7 & 0.10 & 0.33 & 0.01 & 0.06 & 0.03 \\
... & ... & ... & ... & ... & ... & ... & ... & ... & ...  \\
\end{longtable}
\end{center}
%

\section{Cluster Effects on Planets}
\label{sec:planets}
\begin{figure}
\begin{center}
\includegraphics[width=0.9\textwidth]{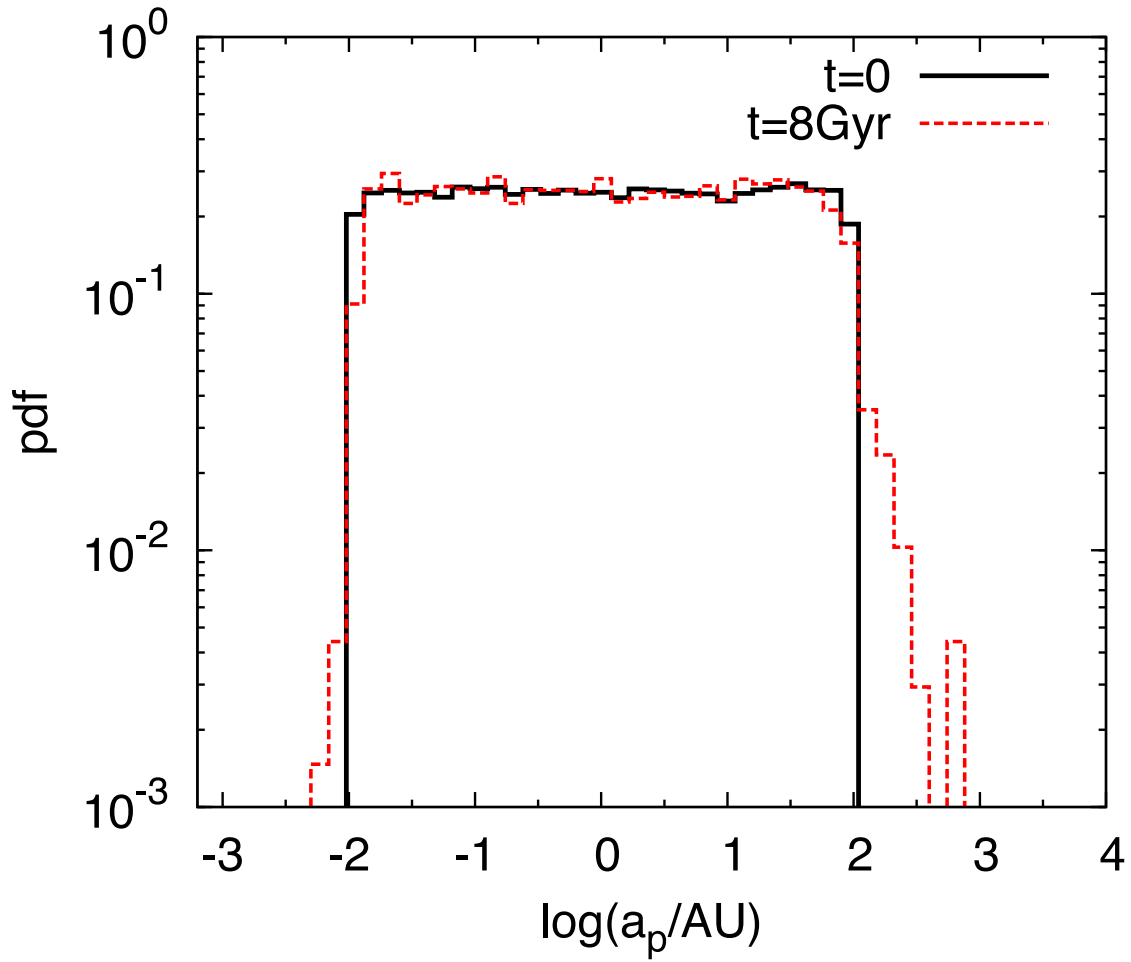}
\caption[$a_p$-distribution of planets]{Initial (black, solid) and final ($t=8\,\gyr$; red, dashed) 
probability density function of the planetary semimajor axis for all planets ($a_p$) in the cluster.  
Note that the fraction of planetary orbits with very small $a_p\leq0.02\,\au$ gets reduced.  
This is mainly due to some stars evolving and engulfing close-in planets, and shrinking of orbits due to tides. 
A small fraction of large $a_p$-orbits expand via interactions as well as stellar evolution driven 
mass loss.  Nevertheless, the shape of the distribution for most orbits is unchanged indicating 
low rate of interactions.  
  }
\label{plot:adist}
\end{center}
\end{figure} 

Here we investigate the effects of stellar encounters on planetary orbits as a function of the planetary 
semimajor axis $a_p$ in open clusters similar to NGC~6791.  

\subsection{Stellar encounter and planetary orbits}
\label{sec:encounter_orbit}
Figure\ \ref{plot:adist} shows the $a_p$-distributions of the planetary objects in our numerical 
model of NGC~6791 (Table\ \ref{tab:NGC6791}) at $t = 0$ and 
$8\,\gyr$.  The $a_p$-distribution at $8\,\gyr$ remains very similar to the initial distribution 
apart from a moderate spreading of the $a_p$ ranges.  For $a_p\leq0.02\,\au$ the planetary 
orbits can shrink due to tidal 
damping.  Moreover, 
stellar encounters make hard planetary orbits harder.  Similarly, soft orbits expand via 
stellar encounters.  In addition, stellar-evolution driven mass loss from the host star also 
expands planetary orbits.  
However, the overall shape of the $a_p$-distribution remains more or less unchanged 
indicating that stellar encounters in this cluster do not significantly change most of the planetary 
semimajor axes with $a_p\leq100\,\au$.  

\begin{figure}
\begin{center}
\includegraphics[width=0.9\textwidth]{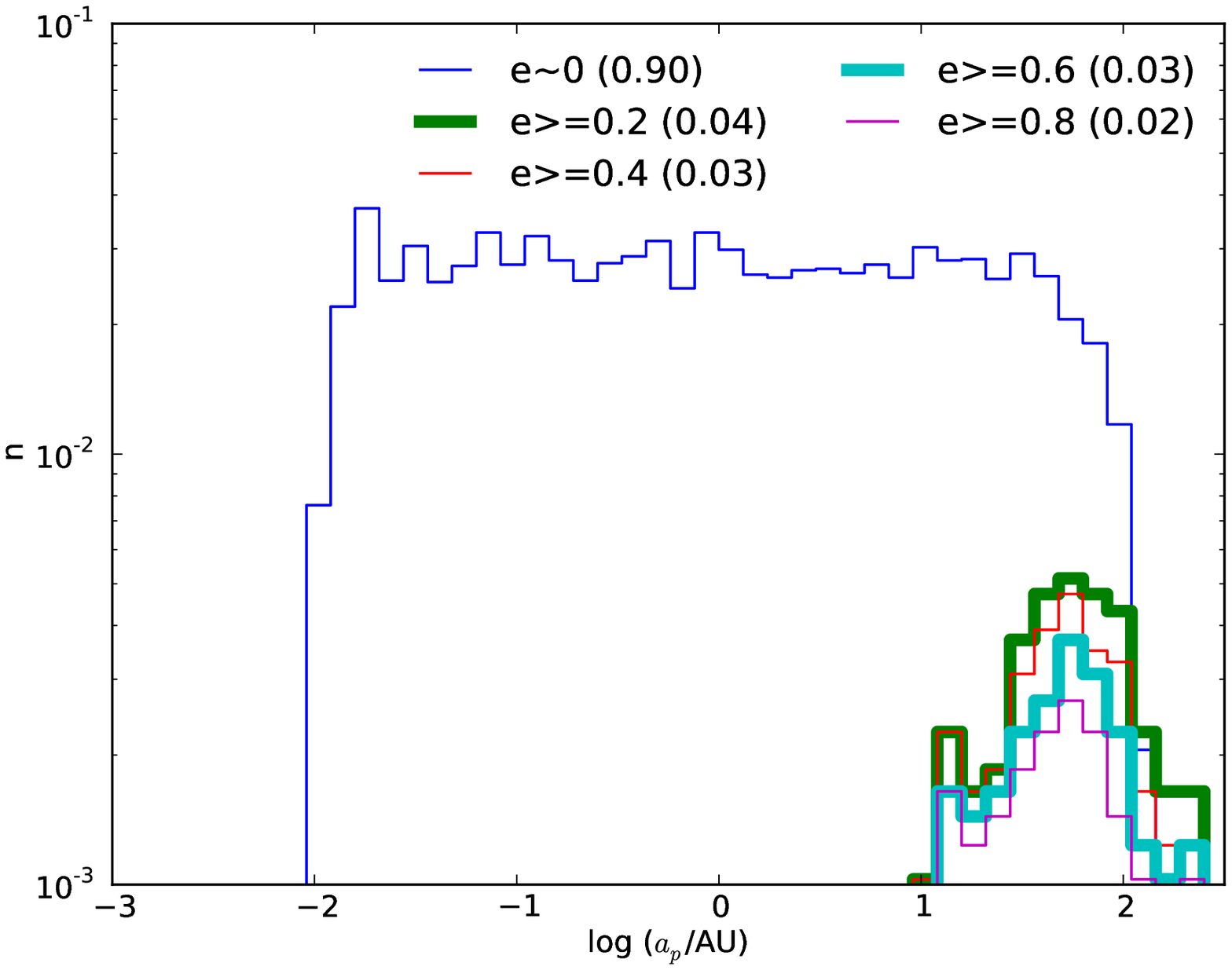}
\caption[$e$-distribution vs $a$]{Semimajor axis distributions of all planets with 
$e\geq$ the given value.  Since all planet orbits are initially circular 
a non-zero $e$ denotes that the planet have interacted via stellar fly-by, or more complicated 
binary-binary interactions.  Close orbits, detectable by transit searches remain 
mostly unperturbed in a cluster like NGC~6791.  Only planets with large-$a_p$ 
($a_p \geq 10\,\au$) orbits have a significant chance of being perturbed.
The fraction of planetary orbits contained inside each histogram is listed in parentheses.   
Only a small ($\sim 10\%$) fraction of planetary orbits gain $e$ via encounters.               
  }
\label{plot:aedist}
\end{center}
\end{figure} 
A more direct way to probe the importance or the lack thereof 
of stellar encounters in changing the planetary orbits in clusters similar to NGC~6791 is 
investigating the number of planets with non-zero eccentricities as a function of $a_p$.  
We initialize all planets on circular 
orbits and all non-zero eccentricities are results of stellar encounters.  Figure\ \ref{plot:aedist} 
shows the $a_p$-distribution of planet orbits with $e\ge$ some given value.  Overall, only 
about $10\%$ of all planetary orbits acquire $e>0.1$ via interactions.  Furthermore, 
the close-in orbits ($a_p < 10\,\au$) are mostly unperturbed.  
Since transit surveys preferentially find planets with small semimajor axes, 
transit-detectable orbits are expected to remain unchanged against possible 
stellar encounters in clusters like NGC~6791.  
Thus, these orbits are {\em not} expected to be different from 
those in the field.  We find the same general result for all our simulated models in {\tt Set1} although 
the exact fraction of significantly perturbed planets depends on the stellar density of the 
clusters (see Table\ \ref{tab:list}).    

Our simulations include only single planet systems.  In multi-planet systems 
indirect instabilities may occur when a stellar encounter changes the outer planet's orbit and 
this excitation in turn 
excites the orbits of the inner planetary system \citep[e.g.,][]{2004AJ....128..869Z,2012arXiv1204.5187B}.  
Nevertheless, the low encounter rates even for large-$a_p$ orbits ($a_p \geq 10\,\au$) 
indicate that such indirect instabilities should be rare and limited to a 
few percent effect in clusters similar to NGC~6791 (Figure\ \ref{plot:aedist}).  
Of course, independent of stellar encounters, planet-planet 
interactions in multi-planet systems can perturb planetary orbits \citep[e.g.,][]{2008ApJ...686..580C,2011Natur.473..187N}.  
Indeed, planet-planet interactions is a common source of perturbations for planetary systems 
orbiting field stars and cluster stars and the excitations due to stellar encounters serve as an 
additional source of perturbation in star clusters.  Our simulations suggest that for clusters 
similar to NGC~6791, perturbations from stellar encounters should be a sufficiently small effect 
so that planets detectable via the transit method should not be much different from those 
in the field.        

Even the very long-period planets ($a \approx 10$ -- $100\,\au$) mostly remain bound to 
the host stars for the whole lifetime of clusters similar to NGC~6791 (Figure\ \ref{plot:adist}).  
The largest orbits have the highest interaction cross-section.  These orbits are 
dynamically soft, resulting in further expansion of these orbits via typical stellar 
encounters in these clusters. In addition, stellar evolution driven mass loss further expands some 
of these orbits.  Our simulations suggest that a small fraction of planetary 
orbits can expand to $a_p \sim1000\,\au$ and still remain bound to their hosts.  
When such clusters dissolve near the Sun, planets with these large-$a$ orbits would 
populate the field and provide potential targets for direct imaging planet searches while 
the planets are still young and luminous.

\begin{figure}
\begin{center}
\includegraphics[width=0.9\textwidth]{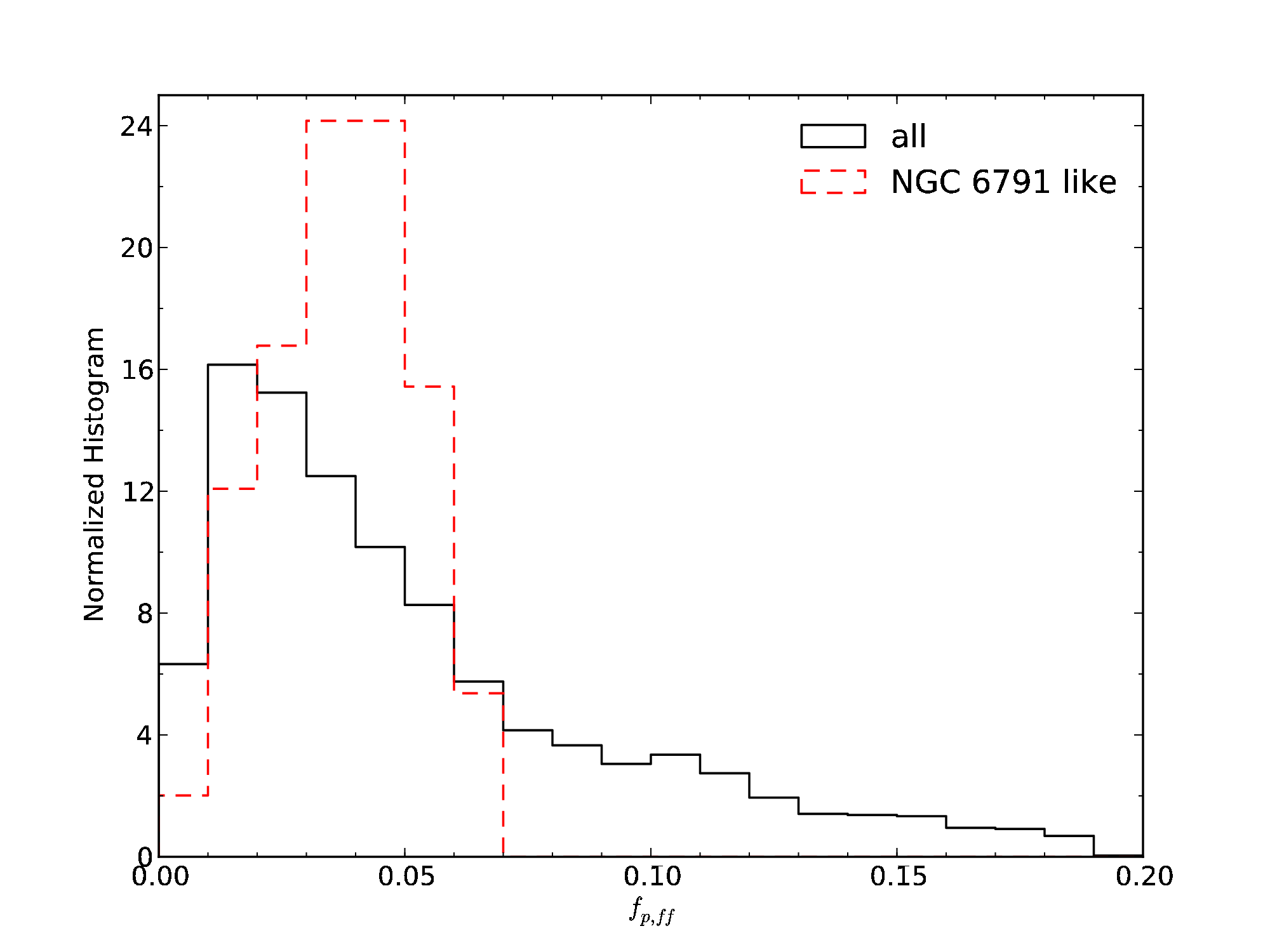}
\caption[Fraction of free-floating planets]{Histogram for the 
fraction of free-floating planets ($f_{p, ff}$, normalized so that the integrated area under the curve is unity) 
for our collection of simulated cluster models 
$f_{p,ff} \equiv (N_{p, ff, bound} + N_{p, ff, esc})/N_{p}(a_p>10\,\rm{AU}, t=0)$, where $N_{p, ff}$ 
is the number of free-floating planets still bound to the cluster potential, $N_{p, ff, esc}$ is 
the number of free-floating planets escaped from the cluster potential via ejections and tidal 
truncation due to Galactic tides, and $N_{p}(a_p>10\,\rm{AU}, t=0)$ is the initial number of 
planets with wide ($a>10\,\rm{AU}$) orbits.  The solid (black) and dashed (red) lines are for all models 
and models similar in properties with those of the NGC~6791, respectively.  
The data from all simulations are listed in Table\ \ref{tab:list}.        
}
\label{plot:free-floating}
\end{center}
\end{figure} 
\subsection{Free-floating planets}
\label{sec:free_floating}
\begin{figure}
\begin{center}
\includegraphics[width=0.9\textwidth]{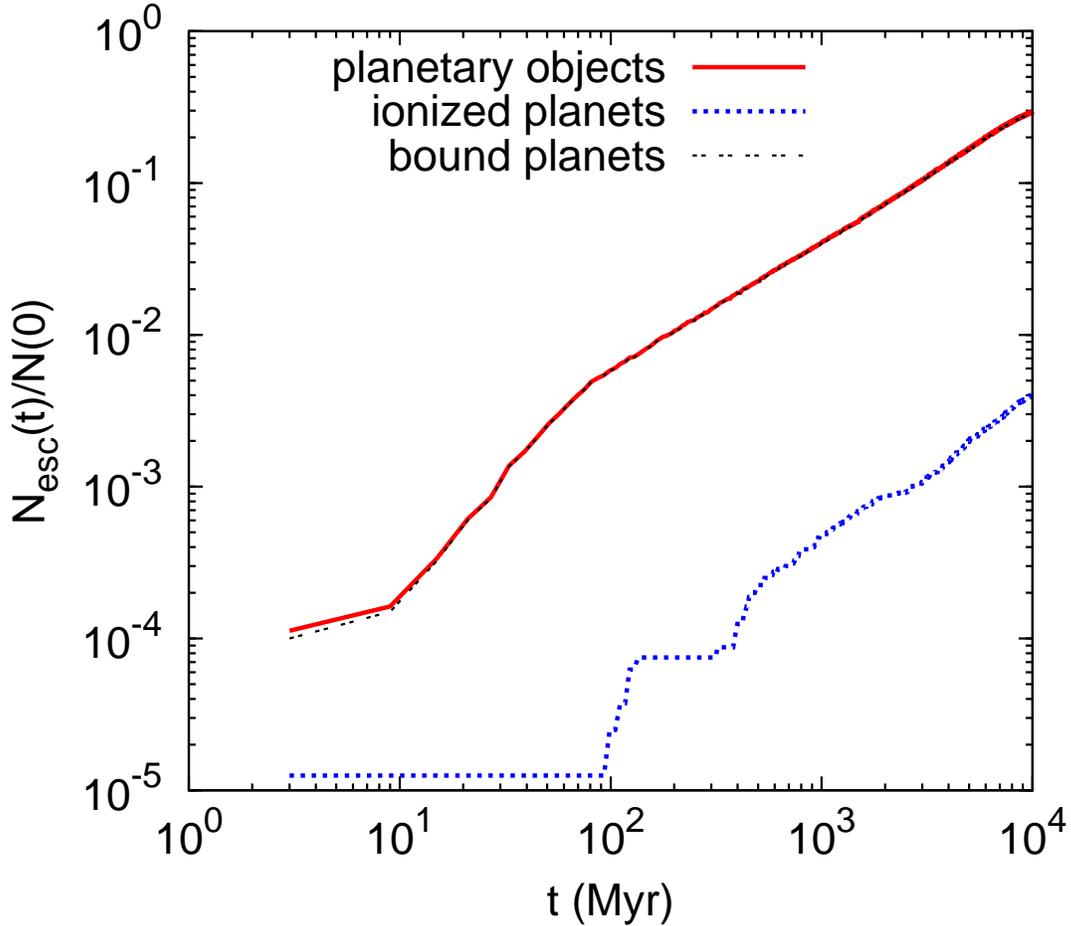}
\caption[Ejected objects from the cluster]{Cumulative fraction of planetary objects leaving the 
cluster potential as a function of cluster age.  Planets may leave the cluster 
due to dynamical ejections resulting from strong binary-mediated 
interactions predominantly in the core of the cluster or due to Galactic tidal stripping 
from the tidal boundary of the cluster as a function of cluster age.  Solid (red), dashed (black), 
and dotted (blue) lines denote all planetary objects, planetary objects still bound to a 
host star, and host less planetary objects, respectively.  All numbers are normalized by the 
initial number of planets $N(0)$.  Note that only a few planets lose their host stars due to stellar 
interactions in NGC~6791.  Most planets are lost from the cluster as the planets' hosts 
escape from the cluster potential.    
  }
\label{plot:escapers}
\end{center}
\end{figure} 

Only a small fraction of 
planetary orbits are significantly perturbed (Figure\ \ref{plot:adist}).  Most of these planets 
initially have large-$a_p$ ($>10\,\au$) orbits (Figure\ \ref{plot:aedist}).  
A fraction of these large-$a_p$ orbits can be disrupted via stellar encounters creating 
free-floating planets.  
Mass segregation drives the free-floating planets away from the cluster center, but the 
free-floating planets can 
remain bound to the cluster potential for many relaxation times near the outskirts 
of the cluster \citep{2002ApJ...565.1251H}.  
In our simulations we find that these free-floating planets reside 
near the tidal boundary of the cluster confirming previous results.  Some free-floating planets 
can remain bound to the cluster for a few billion years after formation.  Since the free-floating 
planets leave the core of the cluster due to mass segregation quickly, further interaction of 
free-floating planets with other planetary orbits is unlikely.  

The number of free-floating planets generated by stellar encounters remains 
low ($\leq 100$), a direct effect of low encounter rates in these clusters.  
This number depends on the actual number of planets with sufficiently wide 
orbits for interaction with other stars or binaries in the cluster.  
Hence we define a quantity $f_{p, ff, bound} \equiv N_{p, ff, bound}/N_p(a>10\,\au)$, where $N_p$ is the number 
of planets bound to hosts in the cluster, $N_p(a>10\,\au)$ the same but with orbits wider than $10\,\au$, 
and $N_{p, ff, bound}$ is the number of free-floating planets bound to the 
cluster potential.  For clusters similar to NGC~6791 $f_{p, ff, bound}$ 
remains a few percent for all times.  In denser clusters in our simulations 
$f_{p, ff, bound}$ can be as high as $\sim20\%$.  At $t_{cl} = 0$ $f_{p, ff, bound} = 0$ by construction.  
In general, $f_{p, ff, bound}$ grows until an age $t_{cl} \sim 1\,\gyr$ followed by a steady decrease 
as the cluster grows older due to preferential stripping of free-floating planets via Galactic tides (Table\ \ref{tab:list}).  

Microlensing surveys indicate that there is a large number of free-floating (or wide orbit) 
planets in the Galaxy \citep{2011Natur.473..349S}.  Microlensing surveys have also detected 
planetary mass candidates in the young open clusters $\sigma$ Orionis and Orion 
\citep{2000Sci...290..103Z,2002ApJ...578..536Z,2009A&A...506.1169B}.  Hence, it is interesting 
to investigate what fraction of wide planetary orbits are disrupted to create free-floating planets that 
may or may not remain bound to the cluster.  For this purpose we define a quantity 
$f_{p, ff} \equiv N_{p, ff}/N_{p, i} (a_p > 10\,\au)$, where $N_{p, ff}$ is the total number of free-floating 
planets (bound to and escaped from the cluster), and $N_{p,i} (a_p > 10\,\au)$ is the initial 
number of wide planetary orbits. 
Figure\ \ref{plot:free-floating} shows the distribution for $f_{p, ff}$ for all our simulated cluster 
models and for all cluster ages.  The values of $f_{p, ff}$ are typically about a few percent for 
our simulated models.  Thus, if stellar encounters dominate production of free-floating planets then 
most microlensing planets are likely wide orbit planets, rather than free-floating planets.  However, 
note that this is a lower limit for $f_{p, ff}$ for clusters similar to our simulated models.  
The actual number of free-floating planets could be significantly higher due to the contribution from planet-planet 
scattering in multi-planet systems \citep[e.g.,][]{2008ApJ...686..580C}.

\subsection{Number of planets and cluster age}
\label{sec:age_number}
The total number of star systems (single and stellar or planetary binaries) bound to the cluster 
steadily decreases over time due to 
Galactic tidal stripping.  As a result the total number of planets in the cluster also steadily decreases over 
time.  However, the ratio of planet host stars in the cluster 
to all star systems in the cluster ($f_p = N_p/N$) remains roughly constant (Table\ \ref{tab:list}).  
For dissolving clusters the ratio 
$f_{p, esc} \equiv N_{p, esc}/N_p$ grows with time, where, $N_{p, esc}$ is the number of planets escaped from 
the cluster potential.  In our simulations, $f_{p, esc}$ can become very large depending 
on the cluster age and properties ($\sim 100$ e.g., {\bf run145}).  
For a completely dissolved cluster the limiting value 
is $f_{p, esc}\rightarrow\infty$.  For our best-match model of NGC~6791 
$f_{p, esc} = 4$ at a cluster age $t_{cl} = 8\,\gyr$ ({\bf run167}, Table\ \ref{tab:list}).  
Thus, for every planet in this cluster 
there are $4$ planets that are now in the field due to the slow dissolution of the cluster.   
   
This decrease in $N_p$ is dominated by the host being lost from the 
cluster.  The number of planets lost from the cluster still bound to their host stars is 
typically $\geq 10^2$ times the number of free-floating planets lost from the cluster at all cluster 
ages (Figure\ \ref{plot:escapers}).  This is simply because complete ionization of planetary 
orbits is not common in clusters such as NGC~6791 even for the broad range in $a_p$ considered 
in our simulations in {\tt Set1}.  The fraction of free-floating escaped planets with respect to escaped planets 
still bound to their hosts ($f_{p, ff, esc}$) are given in Table\ \ref{tab:list} 
for all our simulated clusters at various ages.  
Note again that the fraction of free-floating planets in the field presented here is a lower limit 
based solely on the stellar interactions in the cluster of origin for the planet-host stars and does not 
account for possible contribution from planet-planet scattering.     

\section{Detectability of planets in NGC~6791 using Kepler}
\label{sec:detect}
We have established that stellar dynamics in clusters similar to NGC~6791 alone 
cannot significantly alter small-$a_p$ planetary orbits that are detectable via 
transit surveys like Kepler.  In addition, we know that NGC~6791, already 
in the Kepler FOV, has super-solar metallicity \citep{2011ApJ...733L...1P}.  Hence, formation 
of planets should not be reduced due to lack of solids in protoplanetary disks in this cluster.  
Now we use the simulations from {\tt Set2} (Section\ \ref{sec:models}) to estimate how many 
planets Kepler could discover in NGC~6791 assuming that planets form around cluster stars at the 
same frequency as observed in the field by Kepler and that their initial planetary orbits have similar 
properties to those orbiting around field stars.  
Note that initially we do not include any boost in planet 
occurrence rate due to the super-solar metallicity of NGC~6791 \citep{2005ApJ...622.1102F}.  We discuss 
briefly the estimated effects of the high metallicity of NGC~6791 on the predicted number of planet detections 
later in Section\ \ref{sec:metallicity}.  
The initial conditions for our models in {\tt Set2} 
are taken from the best-match model in {\tt Set1} (Table\ \ref{tab:NGC6791}).  
{\tt Set2} consists of $4$ different realizations of the same initial cluster.  
The period distribution of the planets follow the distribution seen in the Kepler planet candidate list 
after de-biasing for transit probability \citep{2012arXiv1202.5852B}.  
Initially we consider a planet mass distribution estimated from the RV surveys.  
However, later in this section (Section\ \ref{sec:metallicity}) we explore 
how the results would change if we adopt other planet size distributions including ones based on 
Kepler observations.  The $f_p$ is 
chosen based on the observed field planet occurrence rate (Section\ \ref{sec:modeling_planet}).        

\begin{figure}
\begin{center}
\includegraphics[width=0.9\textwidth]{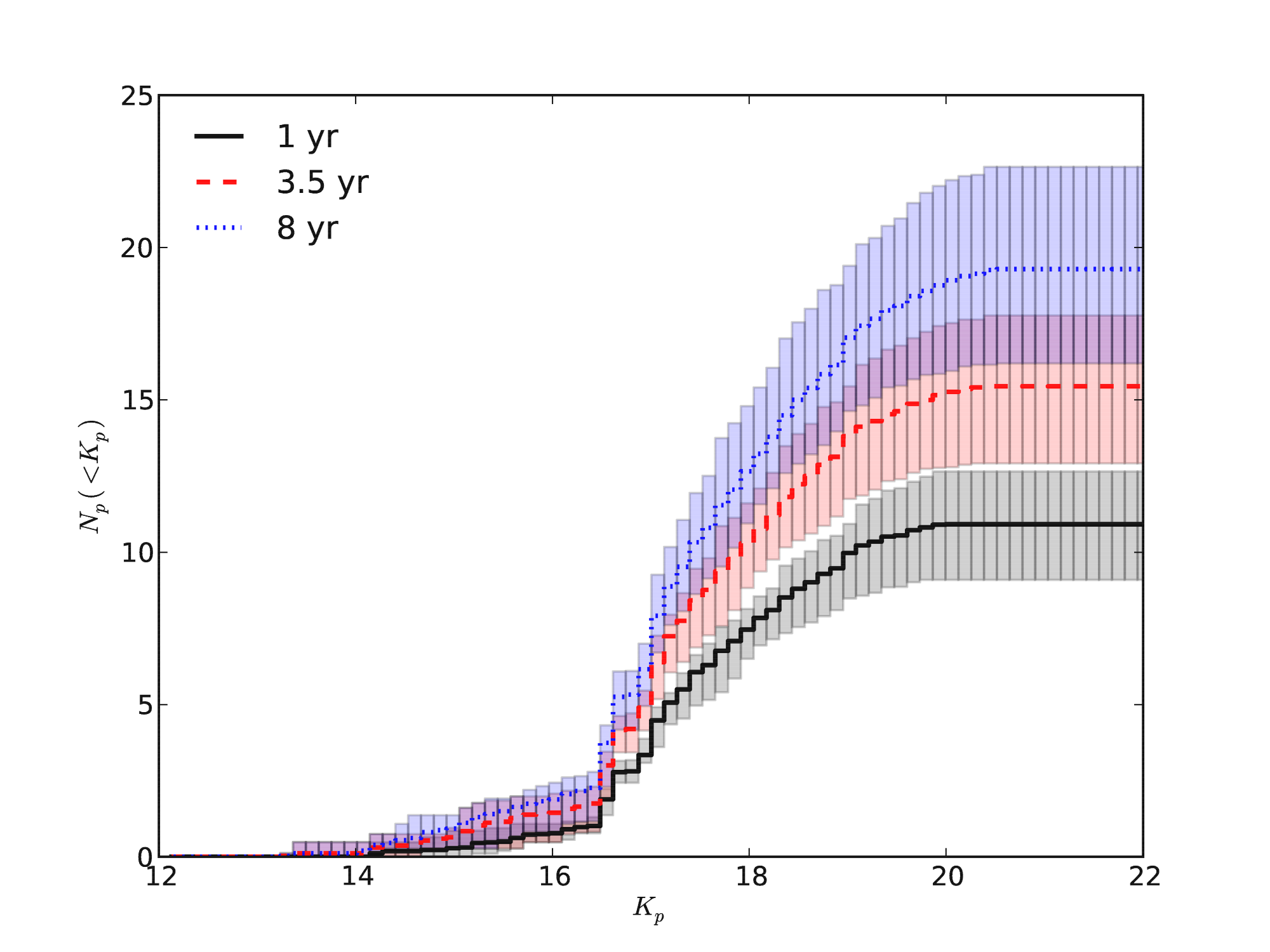}
\caption[$N_p(<K_p)$ vs $K_p$]{Cumulative histogram of the number of Kepler-detectable ($SNR > 7$) 
transiting planets, $N_p(<K_p)$, as a function of $K_p$.  Solid (black), dashed (red), and dotted (blue) 
lines assume Kepler missions extending for $1$, $3.5$, and $8\,\yr$, respectively.  
The black, red, and blue shaded regions denote the range of $N_p(<K_p)$ between all 
$4$ realizations of simulations using the same initial conditions indicating statistical fluctuations.  
Note that $N_p(<K_p)$ rises sharply near the MS turn-off for NGC~6791 at $K_p = 16.5$ (Figure\ \ref{plot:hrd}).  
About $10$ planets are expected to be detectable by Kepler 
by analyzing just $1\,\yr$ of Kepler data if relatively fainter stars ($16<K_p<20$) 
are also fully analyzed. If an extended mission provides $8$ years of data, then the number of 
detected planets could grow to about $20$ including several smaller planets.  Note that these numbers are dependent 
on the overall normalization $f_p$ and also the adopted planet-size distribution (see text; Table\ \ref{tab:np_rpdist}).     
  }
\label{plot:n_kp}
\end{center}
\end{figure} 

Figure\ \ref{plot:n_kp} shows a cumulative histogram of the number of transiting planets ($N_{p} (<K_p)$) 
detectable by Kepler as a function of the Kepler magnitude.  
$K_p = 16.5$ is near the 
MS turn-off of NGC~6791 (Figure\ \ref{plot:hrd}) and the steep rise in $N_{p}(<K_p)$ near 
$K_p = 16.5$ is reflective of the stellar properties in the cluster.  
Although the giants ($K_p < 16.5$) are much brighter than the MS stars, they are less numerous.   
In addition, the giants also have a much larger 
$R_\star$.  Hence the transit $SNR$ is not sufficient for most planets to be detected 
around these giant stars.  However, there is a chance that Kepler may detect a few giant planets around 
low-luminosity giant stars.  Most planet detections in NGC~6791 with Kepler will be around high-luminosity 
MS star hosts ($20 > K_p > 16.5$).  
Our results suggest that if planet occurrence rate in NGC~6791 is similar 
to that observed in the field, Kepler could detect a few to $10$ planets depending on the adopted 
planet-size distribution even with a single year of data collection.  
A total of about $3$ -- $15$ and 
$4$ -- $20$ planets are expected to be detected from $3.5\,\yr$ and $8\,\yr$ of data collection by Kepler.  These 
ranges reflect uncertainties from the planet-size distribution (Table\ \ref{tab:np_rpdist}).  
Relatively faint stars ($K_p < 20$) must also be analyzed to achieve this yield.  
The lack of detections of Kepler candidates in this cluster to date 
may be due to inadequate analysis of these faint target stars.    

\begin{figure}
\begin{center}
\includegraphics[width=0.9\textwidth]{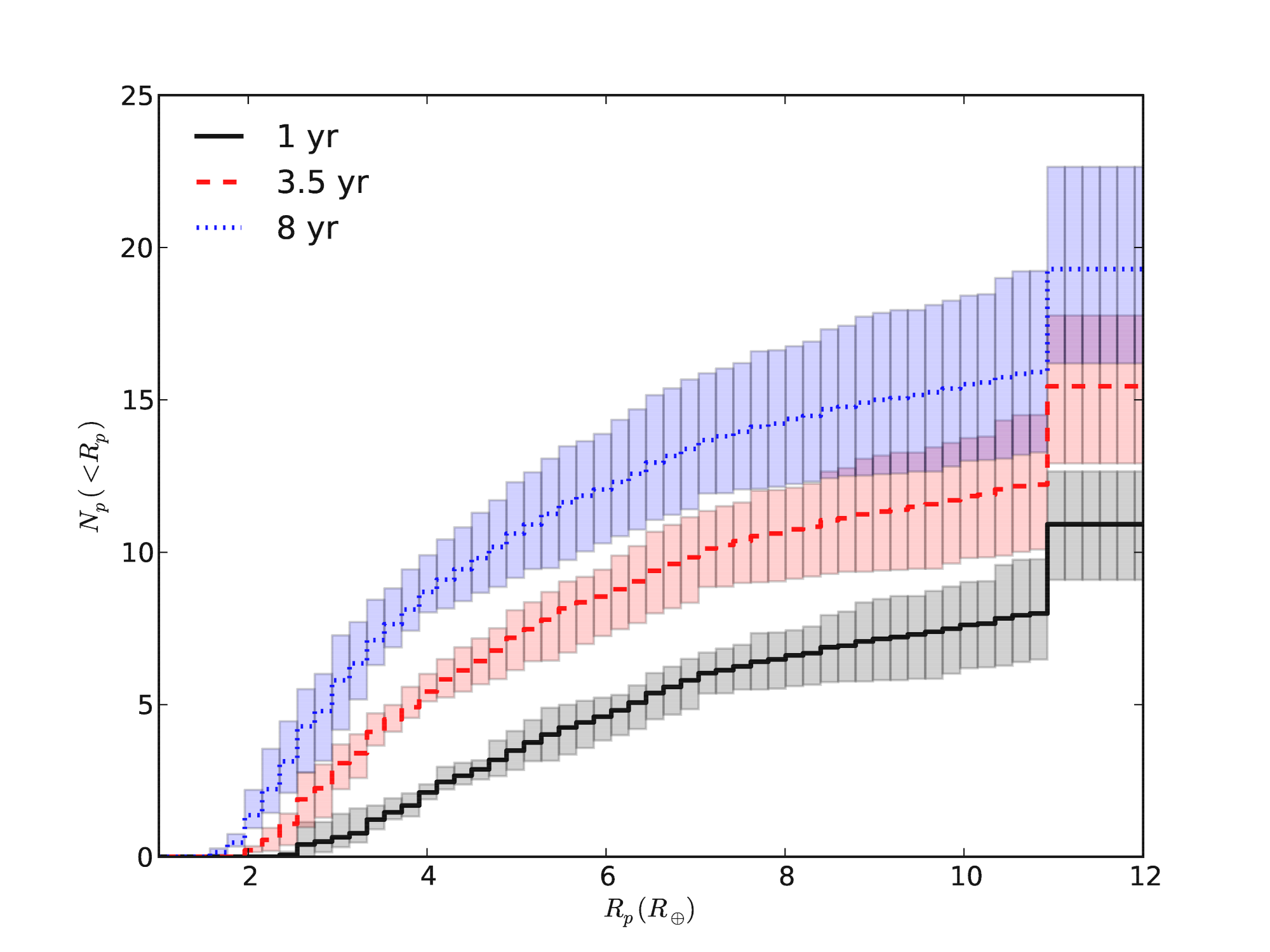}
\caption[$n(<R_p)$ vs $R_p$]{Cumulative histogram of the number of Kepler-detectable ($SNR > 7$) 
transiting planets as a function of the 
planetary radius.  The lines, colors and shades have the same meaning as in Figure\ \ref{plot:n_kp}.  
For observations over one year, the median of the radii of detected planets 
is $R_p \approx 7\,\rearth$. 
An extended mission is expected to detect several smaller planets. 
An extended mission with $8$ years of data collection may result in a handful of planets with $R_p \approx 2\,\rearth$.  
Note that the actual number will vary depending on $f_p$ and the planet-size distribution.    
}
\label{plot:n_rp}
\end{center}
\end{figure} 
Figure\ \ref{plot:n_rp} shows the cumulative number of transiting planets detectable by Kepler in NGC~6791 
as a 
function of the planetary radius ($R_p$). About $35\%$ and $85\%$ of 
all Kepler-detectable planets are expected to be larger than Saturn and Neptune, respectively, if 1 year 
of data is analyzed. Increasing the 
duration of analyzed Kepler data increases the number of detections for smaller 
planets. For example, the median $R_p$ for the predicted number of Kepler-detectable planets is 
about $7$, $5$, and $4\,\rearth$ 
when analyzing $1$, $3.5$, and $8\,\yr$ of Kepler data, respectively. The expected numbers of detectable 
planets with $R_p \leq R_{\rm{Neptune}}$ are about $2$, $5$, and $9$ when analyzing $1$, $3.5$, and $8\,\yr$ of 
Kepler data, respectively. Note that the actual numbers of detections may be higher or lower 
by a few planets due to statistical fluctuations as shown by the shaded regions in Figures\ \ref{plot:n_kp} and 
\ref{plot:n_rp}. The sharp 
increase in the detectable planet numbers in the $R_p = R_J$ bin is reflective of our assumed 
mass-to-radius relation which does not allow our planets to have $R_p > R_J$ 
(Section\ \ref{sec:models}).  Without this assumption some of the planets in this bin would spread into 
bins with larger radii. Almost all detectable planets for $1$, $3.5$, and $8\,\yr$ observations 
are expected to be in orbits with periods below about $40$, $80$, and $100$ days, respectively.  

\begin{figure}
\begin{center}
\includegraphics[width=0.9\textwidth]{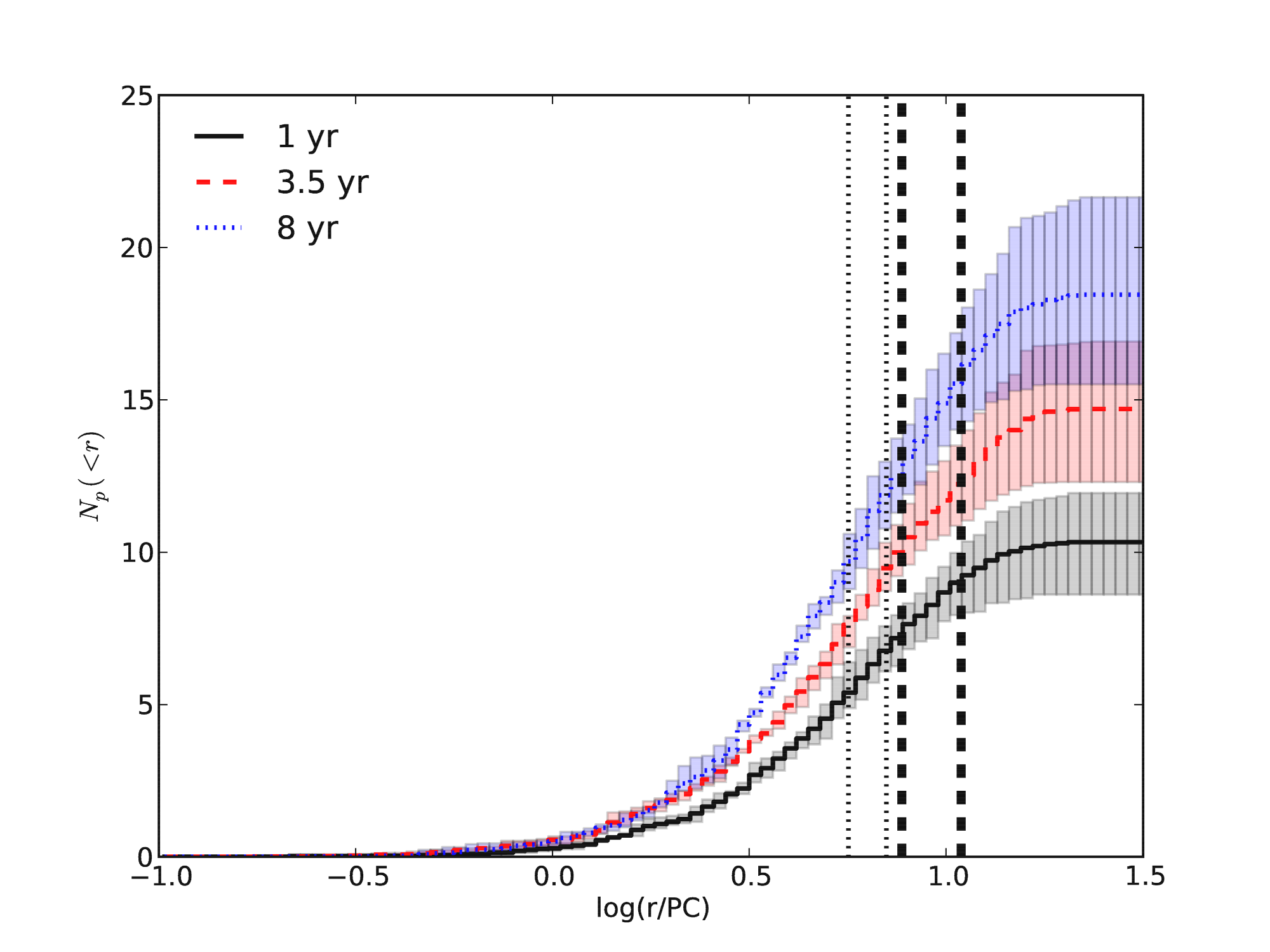}
\caption[$n(<r)$ vs $r$]{Cumulative histogram of the number of the Kepler-detectable 
planets ($SNR>7$) as a function of the planet's sky-projected distance from 
the center of the cluster.  Curves and shades have the same meaning as in Figure\ \ref{plot:n_kp}.  
About $70\%$ of all planets reside outside the core ($r_c = 3.8\,\pc \approx 3.3'$) of the cluster.  
A little over $60\%$ of these planets reside outside the half-light radius ($r_{hl} = 5.1\,\pc \approx 4.4'$) 
of the cluster.  The vertical thin dotted lines mark the projected distances from the cluster center where 
the number of stars within one EE95 for Kepler drops below $1$ assuming an EE95 size of either 
4.5 or 5.6 pixels ($r_{\rm{EE95}} = 5.6, 7\,\pc \approx 4.8', 6'$, respectively).  
Although the actual numbers can vary depending on $f_p$ and the planet-size distribution, the overall 
radial distribution should not change.  The vertical thick dashed lines denote the boundary of 
the $200\times200$ square pixel block centered on the cluster center for the superaperture where all 
pixel level data is being downloaded.  These two lines (for superaperture) denote radial 
distances from the center of the cluster along a side ($6.7'$) and along a diagonal ($9.4'$) of the square 
superaperture box.  For reference, one PC is equivalent to 1.16 arcsecond at the distance 
of NGC~6791 \citep{2011ApJ...733L...1P}.     
}
\label{plot:n_r}
\end{center}
\end{figure} 
Figure \ \ref{plot:n_r} shows the cumulative number of Kepler-detectable planets as 
a function of the sky-projected distance from the center of the cluster.  Most Kepler-detectable 
planets are not in the most concentrated part of the cluster near the center.  
In fact, only about $30\%$ of detectable planets are expected to reside inside the projected 
$r_c = 3.8\,\pc \approx 3.3'$ of the cluster.  A little over $60\%$ of the detectable planets are expected to 
reside outside the projected half-light radius $r_{hl} = 5.1\,\pc \approx 4.4'$ of the cluster.  
This is due to mass segregation 
in the cluster and does not include observational selection effects due to crowding.    

In general, crowding of stars can become a potential problem for transit searches in a star cluster.  
The extent of crowding for each star in the Kepler FOV depends on the number of pixels 
enclosing $95\%$ of the star's flux (EE95).  The EE95 values for the module containing 
NGC~6791 in the Kepler FOV varies between $4.5$ -- $5.6$ pixels (obtained from the Kepler 
Instrument Handbook; http://keplergo.arc.nasa.gov/calibration/KSCI-19033-001.pdf) depending 
on the quarter of observation.  Adopting EE95 of $6$ pixels and an angular scale per pixel 
of $3.98$ arcseconds 
\citep[equivalent to an area of about $0.2\,\pc^2$ at the distance of NGC~6791;][]{2010ApJ...713L..97B}, 
the number density of stars with $K_p < 22$ in NGC~6791 is $<1$ at a cluster-centric two dimensional ($2D$) 
projected distance ($r_{\rm{EE95}}$) of about $7\,\pc$ ($\approx 5.6\,\pc$ if EE95 is $4.5$ pixels; Figure\ \ref{plot:n_r}).  
Between $33$ -- $50\%$ of all otherwise detectable planets 
in our simulations reside outside $r_{\rm{EE95}} = 5.6$ -- $7\,pc \approx 4.8'$ -- $6'$.  
Even at the center of NGC~6791 the number of stars per pixel is $<1$.               
Hence, crowding should 
not be a fundamental limit for planet search in NGC~6791 using Kepler.  However, 
the current pipeline may not be sufficient due to technical issues associated with targeting and 
analysis of stars in a dense field.     

The above estimated numbers of Kepler-detectable planets in NGC~6791 
can change based on various parameters including the intrinsic planet frequency, the planet-size distribution, 
metallicity of NGC~6791, limited data availability and crowding.  
The effect of the assumed intrinsic planet-occurrence frequency, denoted by $f_p$ in this study, 
on the number of detectable transits is straightforward.  If all other parameters such as the period distribution and 
the planet-size distribution are kept fixed, the number of transit detections by Kepler simply scales linearly 
with $f_p$.  In the following sections we discuss the effects of the other parameters on the number of 
detectable transits by Kepler in NGC~6791.  

\subsection{Effects of planet-size distribution}
\label{sec:planet-size}
For a fixed $f_p$, estimating the change in the number of Kepler detectable planets depending on the 
distribution of planet sizes is relatively more complicated.  Assuming that the planet size distribution is 
described by a simple power-law of the form $df/dR_p \propto R^\alpha_p$, we estimate the dependence 
of the number of expected transit detections by Kepler in NGC~6791 as a function of the power-law exponent 
$\alpha$ in the following way.  We extract the planet-harboring stars in our simulations at $8\,\gyr$ but now 
change the planetary sizes based on several different power-law distributions.  For each value of $\alpha$, 
$200$ realizations are generated to estimate the statistical fluctuations.  Each of these planetary systems 
are then analyzed using the same method described in Section\ \ref{sec:modeling_signal}.  
We use a range of $\alpha$ values spanning those reported in literature.  
We report the expected number of transit detections in NGC~6791 by Kepler as a function of $\alpha$ in 
Table\ \ref{tab:np_rpdist}.  Depending on the value of $\alpha$, the expected number of detections can have 
values over an order of magnitude.    

%
\begin{center}
\begin{longtable}{ccccc}
\caption{The expected number of Kepler-detectable transiting planets as a function of $\alpha$, where 
$\alpha$ is the power-law exponent for an assumed planetary size distribution of the form 
$df/dR_p \propto R_p^\alpha$.  Sources (where exist) for $\alpha$ values are given 
in the ``Comments" column.  These estimates do not include any boost of planet occurrence due to the 
super-solar metallicity of NGC~6791.  These estimates assume that data from the full cluster is available.  } \label{tab:np_rpdist} \\   
\hline \hline \\
\multicolumn{1}{c}{\textbf{$\alpha$}} & 
\multicolumn{3}{c}{\textbf{Number of Detected Transits}} & 
\multicolumn{1}{c}{\textbf{Comments}} \\ \cline{2-4}
\multicolumn{1}{c}{} &
\multicolumn{1}{c}{\textbf{1 year}} &
\multicolumn{1}{c}{\textbf{3.5 year}} &
\multicolumn{1}{c}{\textbf{8 year}} &
\multicolumn{1}{c}{\textbf{}} \\
\hline \hline\\
\endhead
%

\endfoot

$-1.5$ & $12\pm2$ & $16\pm2$ & $19\pm2$ & - \\
$-1.99$ & $11\pm2$ & $15\pm3$ & $19\pm4$ & these simulations; \citet{2010Sci...330..653H} \\
$-2.7$ & $3\pm2$ & $6\pm2$ & $8\pm2$ & \citet{2011ApJ...742...38Y} \\
$-2.92$ & $2\pm1$ & $4\pm2$ & $7\pm2$ & \citet{2012ApJS..201...15H} \\
$-3.5$ & $1\pm1$ & $3\pm2$ & $4\pm2$ & - \\
\hline \hline
\end{longtable}
\end{center}
%

\subsection{Effects of metallicity}
\label{sec:metallicity}
The planet occurrence rate varies with the metallicity of the target stars \citep{2005ApJ...622.1102F}.  
NGC~6791 has 
a super-solar metallicity ([Fe/H] = +0.30; \citealt{2009AJ....137.4949B}).  Analysis up to this point did not 
account for the fact that the high metallicity of NGC~6791 is likely to boost planet occurrence in this cluster 
relative to the Kepler target list.  The correlation between planet occurrence and metallicity is well 
established for giant planets \citep[e.g.,][]{2005ApJ...622.1102F,2008PhST..130a4003V,2011A&A...526A.112S} 
but is not as strong for planets smaller than $4\,\mearth$ 
\citep[e.g.,][]{2011A&A...533A.141S,2012Natur.486..375B}.
Here we explore how much the super-solar metallicity of NGC~6791 could change the estimated number of 
detectable transiting planets by Kepler in this cluster.  
To remain conservative we assume that the planet frequency is increased due to metallicity only for planets 
larger than Saturn.  
We further assume that the distributions of orbital properties (e.g., period distribution) 
of hot jupiters are independent of the metallicity.    

Based on the RV observations \citet{2005ApJ...622.1102F} fit for the planet occurrence rate and 
the metallicity of the host stars and find 
\begin{eqnarray}
\label{eq:func1}
f_p ([Fe/H]) &=& f_{p, \odot} \times \left[ \left( \frac{N_{Fe}}{N_{H}} \right) / \left(\frac{N_{Fe}}{N_{H}} \right)_\odot \right]^2 
\end{eqnarray}
where, $N_{x}$ denotes the number of atom x per unit volume, $f_{p, \odot}$ is the planet 
occurrence rate at Solar metallicity, and $f_{p} ([Fe/H])$ is the planet occurrence rate as a function of the 
metallicity of the host star.  We call this Function1.  To estimate the planet occurrence rate in NGC~6791 
we also consider       
\begin{eqnarray}
\label{eq:func2}
f_p ([Fe/H]) &=& f_{p, \odot} \times \left[ \left( \frac{N_{Fe}}{N_{H}} \right) / \left(\frac{N_{Fe}}{N_{H}} \right)_\odot \right]^2,\,\rm{for~} [Fe/H] \geq 0 \nonumber\\
  & =& f_{p, \odot},\,\rm{for~} [Fe/H] < 0, 
\end{eqnarray}
which we call Function2.  Here we assume that the planet occurrence 
rate with respect to the metallicity increase as a power-law as found by \citet{2005ApJ...622.1102F} 
only for Solar or higher metallicities and is flat below Solar metallicities.  For both cases the average planet occurrence rate 
is calculated considering all Kepler targets using $f_{p, \rm{KIC}} = \frac{1}{n}\sum{f_p ([Fe/H])}$.  The expected planet occurrence 
rate $f_{p, \rm{NGC~6791}}$ is also calculated for NGC~6791 using [Fe/H] = +0.30 \citep{2009AJ....137.4949B}.  The boost of 
planet occurrence rate for giant planets larger than Saturn is then simply $f_{p, \rm{NGC~6791}}/f_{p, \rm{KIC}}$.  We find that this ratio is 
$4.9$ and $3.1$ assuming Function1 and Function 2, respectively.  

If the super-solar metallicity boosts the planet occurrence rate for all planet sizes, then the total number 
of detectable transits will simply scale with the numbers presented in Table\ \ref{tab:np_rpdist}.  
If the planet occurrence rate is boosted by the above factors in NGC~6791 only for giant planets 
(i.e., larger than Saturn), the expected 
number of transit detections will increase by a smaller factor depending on the number of giant planets, 
which in turn depends on the planet-size distribution.    
Table\ \ref{tab:metal} shows the expected numbers of transit detections by Kepler boosted due to the super-solar metallicity of NGC~6791 
assuming that the planet occurrence rate is boosted only for giant planets larger than Saturn and that the 
planet-size distribution is described by a power-law with exponent $\alpha = -1.99$.  
\clearpage
\begin{center}
\begin{longtable}{ccccccccc}
\caption{The expected number of Kepler-detectable transiting planets if the planet occurrence rate in NGC~6791 
is boosted due to its high metallicity.  We consider two different relations relating the giant planet occurrence rate 
and the metallicity of the host star (Equations\ \ref{eq:func1} and\ \ref{eq:func2}) denoted as Function1 and 2 respectively.  
For each relation we estimate the number of expected transit detections for various observation lengths assuming that the occurrence 
of planets more massive than Saturn is boosted due to metallicity.  $t_{\rm{obs}}$ denotes the duration of data collection 
by Kepler.  For each of these data collection durations the number of transit detections are shown both with 
and without the metallicity boost (assuming $\alpha = -1.99$; Table\ \ref{tab:np_rpdist}).  All of the above 
numbers are given for the full cluster (``All Cluster"), the subregion of the cluster where pixel-level data is 
available (superaperture, a $200\times200$ pixels box centered on the cluster center; 
``Downloaded Cluster"; see Section\ \ref{sec:kepler_data}), and the subregion of the cluster within the 
superaperture and outside $r \geq r_{\rm{EE95}}$, where the number of stars per 
EE95 is less than 1 (``$r\geq r_{\rm{EE95}}$").  } \label{tab:metal} \\   
\hline \hline \\
\multicolumn{1}{c}{\textbf{$t_{\rm{obs}}$ (yr)}} & 
\multicolumn{1}{c}{\textbf{Relation}} & 
\multicolumn{2}{c}{\textbf{All Cluster}} & 
\multicolumn{1}{c}{} &
\multicolumn{4}{c}{\textbf{Downloaded Cluster}} \\ \cline{6-9} \cline{3-4}
\multicolumn{1}{c}{} & 
\multicolumn{1}{c}{} &
\multicolumn{1}{c}{} & 
\multicolumn{1}{c}{} & 
\multicolumn{1}{c}{} & 
\multicolumn{1}{c}{} & 
\multicolumn{1}{c}{} & 
\multicolumn{2}{c}{\textbf{$r \geq r_{\rm{EE95}}$}} \\ \cline{8-9}
\multicolumn{1}{c}{} & 
\multicolumn{1}{c}{} & 
\multicolumn{1}{c}{\textbf{Unboosted}} & 
\multicolumn{1}{c}{\textbf{Boosted}} & 
\multicolumn{1}{c}{} &
\multicolumn{1}{c}{\textbf{Unboosted}} & 
\multicolumn{1}{c}{\textbf{Boosted}} & 
\multicolumn{1}{c}{\textbf{Unboosted}} & 
\multicolumn{1}{c}{\textbf{Boosted}} \\
\hline\hline 
\endhead
%

\endfoot
$1$ & Function1 & $10\pm 2$ & $21\pm 7$ &  & $8\pm1$ & $18\pm2$ & $3\pm1$ & $6\pm2$ \\
$3.5$ & (Equation\ \ref{eq:func1}) & $15\pm 2$ &$31\pm 7$ &  & $11\pm1$ & $22\pm2$ & $4\pm2$ & $8\pm2$\\
$8$ &  & $18\pm 3$ & $34\pm 7$ &  & $14\pm2$ & $26\pm3$ & $5\pm2$ & $9\pm2$ \\
\hline
$1$ & Function2 & $10\pm 2$ & $16\pm 5$ &  & $8\pm1$ & $14\pm2$ & $3\pm1$ & $4\pm2$ \\
$3.5$ & (Equation\ \ref{eq:func2}) & $15\pm 2$ & $23\pm 5$ &  & $11\pm1$ & $17\pm2$ & $4\pm2$ & $6\pm2$ \\
$8$ &  & $18\pm 3$ & $26\pm 6$ &  & $14\pm2$ & $20\pm3$ & $5\pm2$ & $6\pm3$ \\
\hline \hline

\end{longtable}
\end{center}
\subsection{Effects for finite data download by Kepler}
\label{sec:kepler_data}
\begin{figure}
\begin{center}
\includegraphics[width=0.9\textwidth]{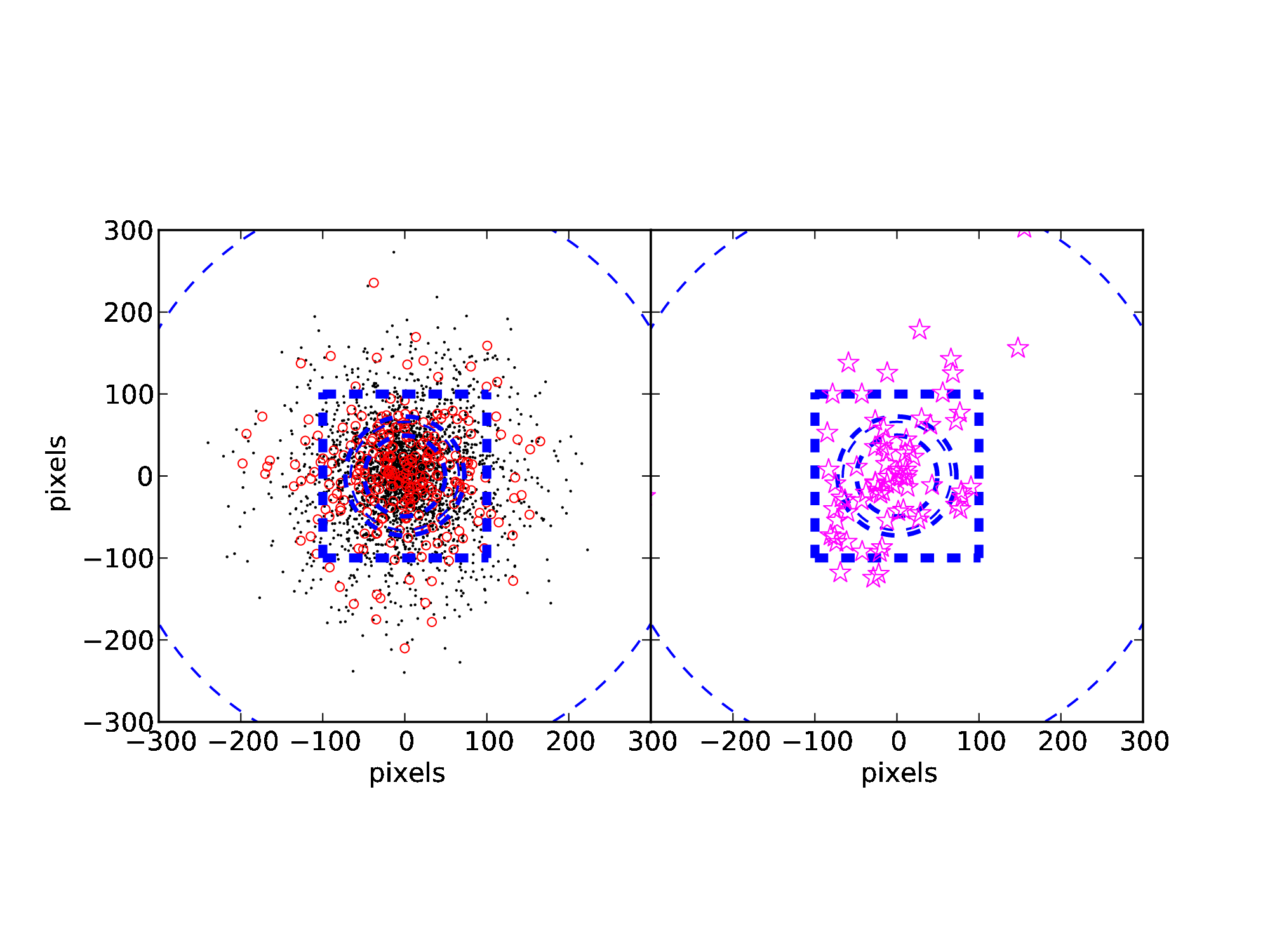}
\caption[2D image of cluster]{Left Panel: Two dimensional image of our simulated model of NGC~6791.  
The grey dots denote all cluster objects (single and binary stars).  The red circles denote planet hosts 
detectable ($SNR > 7$) by Kepler.  Right panel: Magenta stars denote the positions of the individual stars 
for which the positions (within 23 arcseconds from the center of NGC~6791), magnitudes ($K_p>16$), and 
the surface gravities ($\log g>3.5$) are consistent with their being main-sequence stars in NGC~6791, and 
for which data is being (or will be) downloaded by Kepler.  For both panels the dashed square denotes 
the $200\times 200$ pixel block of super apertures where pixel level data is being downloaded.  The 
dashed circles from small to large are the $r_c$, $r_{hl}$, $r_{\rm{EE95}}$, and the tidal radius (partly visible in the scale) for reference.  
Note that the half-light radius is just inside the radius where the number of stars per EE95 is $< 1$.  All lengths are 
denoted in units of Kepler pixels $3.98''$.  
}
\label{plot:2d_image}
\end{center}
\end{figure} 
Our analysis in Section\ \ref{sec:detect} shows that the best chance for Kepler to detect transits in 
NGC~6791 is around bright MS stars ($16.5 < K_p < 20$; Figure\ \ref{plot:n_kp}).  
All predicted numbers of transit detections from our simulations in 
Sections\ \ref{sec:planet-size} -- \ref{sec:metallicity} 
have been calculated assuming that data 
for all stars in NGC~6791 up to $K_p=20$ is available.  In practice, not all needed pixels are 
downloaded.  Kepler has been observing a few regions in the FOV through large, custom apertures where 
pixel level data is downloaded.  These are called ``superapertures" (http://keplergo.arc.nasa.gov/Blog.shtml).  
In particular for NGC~6791 a superaperture of $200\times200$ pixels block centered on the cluster center is 
being downloaded \citep[e.g.,][]{2011ApJ...739...13S}.  
In addition to this superaperture data, data from about $400$ individual stars consistent 
to be MS stars in NGC~6791 (i.e., within the tidal radius $r_t = 23.1'$ of the center of NGC~6791, 
magnitude $K_p > 16$, and surface gravity $\log g > 3.5$) are being downloaded by Kepler.  Most of these individual stars with downloaded data are already included in the superaperture (Figure\ \ref{plot:2d_image}).  
%
%
The right panel of Figure\ \ref{plot:2d_image} shows the superaperture 
block, and the individual stars consistent to be MS stars in NGC~6791 for which data is available 
(or will become available as part of a 
GO proposal).  Important cluster structural radii 
($r_c = 3.3'$, $r_{hl} = 4.4'$, $r_{\rm{EE95}} = 6'$ and $r_t = 23.1'$) are 
also shown.  The left panel shows a two dimensional image of our simulated model of 
NGC~6791 over-plotted with the superaperture 
block, and the above mentioned cluster radii.  Detectable transits by Kepler will 
be around bright main-sequence dwarf stars in NGC~6791.  
Thus at the given age of NGC~6791 these host stars are 
among the lower-mass sub-population of the cluster ($M_\star \leq 1.2\,\msun$).  Mass-segregation distributes these relatively lower mass 
stars at larger cluster centric distances compared to the stellar binaries and more-massive (and evolved) single stars 
in NGC~6791.  Thus while red giants are predominantly near the center of the cluster, a large fraction 
($\approx 60\%$)
of transit detectable planet host stars are outside $r_{hl}$.  Ideally, the best place to search for these 
planets is between $r_{\rm{EE95}}$ and $r_t$.  However, due to the limited data availability this cannot 
be done.  
We find that between $r_{\rm{EE95}} = 4.8'$ -- $6'$ and the superaperture box ($\approx 6.7'$ along a side of the square 
and $9.4'$ along a diagonal) $3\pm 1$ transits could be detectable by 
Kepler with one year of data collection assuming $\alpha = -1.99$ and no metallicity boost.  
Table\ \ref{tab:metal} lists the expected numbers (with and without metallicity boost) of transit detections 
for different observations lengths in different cluster regions (including the region between $r_{\rm{EE95}}$ and 
the superaperture box).    
However, note that most of the stars in the superaperture are not analyzed as part of the regular 
planet search pipeline.

\section{Discussion}
\label{sec:conclusion}
\begin{figure}
\begin{center}
\includegraphics[width=0.9\textwidth]{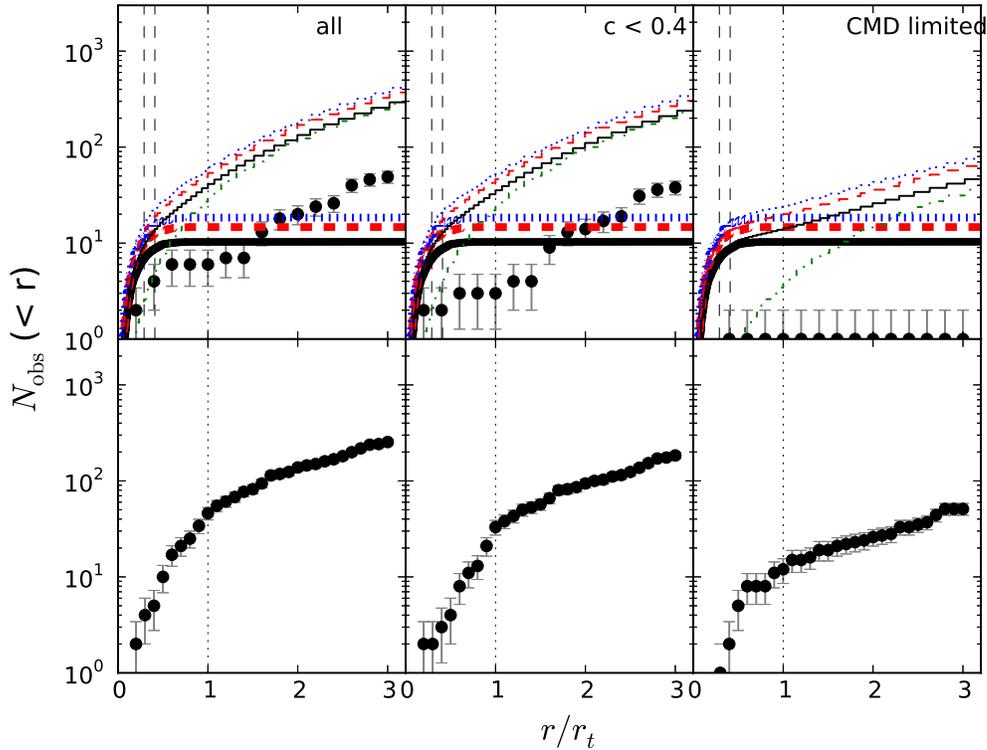}
\caption[]{\footnotesize Cumulative number of observed Kepler candidates $N_{obs}(<r)$ vs the cluster-centric 
positions in units of tidal radii ($r_t$).  Top panels show 
results for the region around NGC~6791 only.  The bottom panels show combined results 
including all four open clusters in the Kepler FOV.  
The left, middle, and right panels include all stars with $K_p<20$ in the Kepler FOV, a subset of 
that with contamination values $c<0.4$, and another subset that are between the single-star 
and equal-mass binary MSs on a $T_{eff}$ vs. $K_p$ diagram for each cluster, respectively.  
Black points show Kepler planet candidates \citep{2012arXiv1202.5852B}. Error bars are standard $1\sigma$ Poisson errors.  
The histograms in each panel show the expected number of planets Kepler could detect based on 
our models.  These histograms are calculated from our models using $\alpha = -1.99$ 
(Section\ \ref{sec:planet-size}), no metallicity 
boost due to the super-solar metallicity of NGC~6791 (Section\ \ref{sec:metallicity}), and assuming 
that data for all stars up to $K_p = 20$ in NGC~6791 is available (Section\ \ref{sec:kepler_data}).  
The thick lines include only detectable transiting planets from our simulated models of NGC~6791.
Thin lines include planets both in our simulated models and those assumed to be around 
foreground/background stars based on the stellar densities inferred from the 
Besan\c{c}on model of the Milky Way \citep{2003A&A...409..523R}.    
The black (solid), red (dashed), and blue (dotted) lines show results based on $1$, $3.5$, and 
$8\,\yr$ observations by Kepler.  The green (dash-dot) lines show results if the cluster had no planets.  
The vertical dotted lines show where the tidal boundary for NGC~6791 is to aid the eye.  The two 
dashed vertical lines in each of the top panels show the positions of the boundaries along a side 
and along a diagonal of the square $200\times 200$ pixels superaperture box centered on the 
center of NGC~6791 (Section\ \ref{sec:kepler_data}; \citealt{2011ApJ...739...13S}).   
}
\label{plot:nposition}
\end{center}
\end{figure} 
We revisit the possibility of detecting planets around normal MS stars in clusters 
in the light of the highly successful Kepler mission.  
In particular, we focus on NGC~6791, an old, massive, and metal-rich open cluster.  
The fact that NGC~6791 is already being observed by Kepler makes it 
an ideal candidate to test hypotheses developed to explain previous planet searches that report a 
dearth of planets around normal MS stars in clusters.  

Using $\sim 200$ detailed numerical simulations of a cluster's evolution 
and planet-harboring stars we study effects of stellar encounters in rich open clusters.  
We choose a model that 
best matches the observed properties of NGC~6791 from this large set of simulations performed 
on a broad multidimensional grid of initial conditions (Section\ \ref{sec:ngc6791}, Table\ \ref{tab:NGC6791}, 
Figure\ \ref{plot:sbp}).  
We find that planetary orbits are rarely disrupted solely via stellar encounters in open clusters 
for a large range in cluster mass and stellar density (Table\ \ref{tab:list}).  For clusters 
similar to NGC~6791, only about 10\% of planetary orbits with relatively large semimajor axes 
($a_p \geq 10\,\au$) are likely to be excited via strong stellar encounters (Figure\ \ref{plot:aedist}).  
A small fraction ($\sim 1\%$) of large-$a_p$ orbits may get ionized (Table\ \ref{tab:list}).  
However, 
the bulk of the planetary orbits, especially the small-$a_p$ orbits ($a_p<1\,\au$), detectable 
via transit searches, remain undisturbed (Figures\ \ref{plot:adist} and \ref{plot:aedist}).  

We find that the number of free-floating planets can range from a few to $\sim 100$ in a 
cluster like NGC~6791 depending on the age of the cluster.  
In our best-match model of NGC~6791 the highest fraction of free-floating planets bound to the cluster potential, 
$f_{p, ff, bound}$, grows from an initial value of zero by 
construction to the maximum value of $0.4$ (Table\ \ref{tab:NGC6791}).   
However, depending on the cluster properties and age $f_{p, ff, bound}$ 
can be higher (Figure\ \ref{plot:free-floating}; Table\ \ref{tab:list}).  
The total number of planets in a cluster steadily decreases with the age of the cluster mainly due to 
Galactic tidal stripping of the host stars (Figure\ \ref{plot:escapers}).  These planet-host stars then 
leave the cluster to populate the field.  Planetary orbits with $a>100\,\au$ (and up to $1000\,\au$ in 
our simulations) can remain bound to the host stars in a cluster similar to NGC~6791 
(Figure\ \ref{plot:adist}).  

\subsection{Prospects for cluster-planet detection from Kepler}
We estimate the expected number of planets Kepler may discover in NGC~6791 using 
the best-match initial conditions obtained from our grid of simulations to create models 
of NGC~6791 with planet frequency and planetary properties (Section\ \ref{sec:modeling_planet}) guided by the 
current Kepler observations in the field \citep{2012arXiv1202.5852B}.  
If the planet occurrence rate in NGC~6791 is the same as observed in the field by Kepler, then 
the existing Kepler data could detect transits of between a few to $10$ planets (Figure\ \ref{plot:n_kp}, Table\ \ref{tab:np_rpdist}) 
depending on the intrinsic distribution for planet sizes analyzing 1 year of data.  
An extended Kepler mission over $8\,\yr$ (or more) is projected to increase this yield by a factor of 2. 
However, due to the old age ($8\pm1\,\gyr$) and 
large distance ($4\,\kpc$) of NGC~6791, most of these detectable-planet-host stars are relatively faint 
($K_p\geq16.5$, the MS turn-off for NGC~6791; Figure\ \ref{plot:hrd}).  Hence, to attain such an 
yield, Kepler data for relatively faint stars ($K_p<20$) must be analyzed properly.  These numbers 
are expected to increase due to the high metallicity of NGC~6791 
\citep[given the positive correlation between the intrinsic planet occurrence rate and the metallicity 
of the host stars, e.g.,][]{2005ApJ...622.1102F} (Table\ \ref{tab:metal}).  On the other hand, since data 
is not available for all stars in NGC~6791 the actual number of transit detection will be reduced 
(Table\ \ref{tab:metal}).       

Although a few giant planets may be 
detectable around low-luminosity giant stars, most planets 
should be detected around MS stars close to the MS turn-off (Figure\ \ref{plot:n_kp}).  
When using only $1\,\yr$ observation by Kepler a large fraction ($35\%$) of the detected planets will 
likely be gas giants.  With longer observation times the expected number of detections for smaller planets 
grows and the median $R_p$ of detected planets decreases from $7\,\rearth$ after $1\,\yr$ to about $4\,\rearth$ 
after $8\,\yr$ (for a planet-size distribution with $\alpha = -1.99$; Table\ \ref{tab:np_rpdist}).  
With an extended mission of $8\,\yr$ a few transits of planets as small as 
$R_p \approx 2\,\rearth$ could be detected in NGC~6791 (Figure\ \ref{plot:n_rp}).    

About $70\%$ of the detectable 
planets are expected to reside outside the core ($r_c \approx 3.3'$) of NGC~6791, 
and a little above $60\%$ of the 
detectable planets are expected to reside outside the half-light radius ($r_{hl} \approx 4.4'$) 
of NGC~6791 (Figure\ \ref{plot:n_r}).  
Between $30$--$50\%$ of all otherwise detectable planets reside outside the $2D$ projected 
distance $r_{\rm{EE95}} = 4.8'$ -- $6'$ 
from the cluster center beyond which the number of stars per EE95 is less than 1.  
Hence, crowding should not be a limiting factor in the detection of planets in NGC~6791 
using Kepler, once the data between 
$r_{\rm{EE95}}$ and the superaperture boundaries ($6.7'$ along a side and $9.4'$ along a diagonal) 
is fully analyzed.  

Note that the MS planet hosts, are among the lower mass ($M_\star \leq 1.2\,\msun$) 
subpopulation in the cluster given the old age of NGC~6791.  
Hence mass segregation distributes them at relatively larger separations 
from the cluster center compared to the stellar binaries and more massive evolved stars 
(Section\ \ref{sec:detect}).  Ideally, the best place to search for transits in this cluster would be between 
$r_{\rm{EE95}}$ (to avoid crowding) and $\approx 0.5r_t$ (since there are simply not many stars 
beyond this radius).  Hence, based on our results we suggest the following.  \\
1. Relatively fainter stars ($K_p < 20$) should also be properly analyzed.  \\
2. Ongoing observations of the clusters in the Kepler FOV, especially the relatively massive clusters 
NGC~6791 and NGC~6819 are maintained not only for astrophysics, but also to search for exoplanets.  \\
3. The pixel-level data obtained from the superapertures are analyzed for existence of transiting planets.  
Especially, data for regions outside $r_{\rm{EE95}}$ where there is reduced crowding.  \\
4. For detection of exoplanets it is more important to get data from the outer (e.g., outside the $r_{hl}$) regions of 
a cluster.  Considering the limitation of the number of pixels that can be downloaded, 
we suggest that some pixels 
from the center of the superapertures be reassigned to search for planets around individual MS stars 
likely to be cluster members and located outside the superaperture block.    

Kepler results can determine whether planets in short period orbits are as common around cluster stars 
as they are around field stars.  Even a null result from Kepler would be interesting since that would potentially 
provide stronger constraints than the existing results on the frequency of planet occurrence around MS stars 
in star clusters. 

\subsection{How likely is a planet discovered near a cluster an actual cluster member?}
In Figure\ \ref{plot:nposition} we show a preliminary analysis of the Kepler planet candidates near 
star clusters as a function of their sky 
positions relative to a cluster.  Planet candidates are from \citep{2012arXiv1202.5852B} 
and are based on searching Kepler data from Q1--Q6.  The top panels show the region around NGC 6791 only, 
and the bottom panels combine the regions around four open clusters 
(NGC~6866, NGC~6811, NGC~6819, and NGC~6791) in the Kepler FOV.  We plot the 
number of observed Kepler candidates ($N_{\rm{obs}}$) as a function of the projected distance from the cluster 
center(s).  
The cluster-centric distances are given in units of the cluster's tidal radius(ii) ($r_t$).  
Cluster parameters, including positions, tidal radii, distances, ages, and redennings are taken from the 
literature \citep[][WEBDA]{2001AJ....122..266K,2005A&A...438.1163K,2009AJ....138..159H,2008A&A...477..165P,2011ApJ...733L...1P}
The observed points are drawn for three sets of target stars defined in the following way.  
In the left panels we include all stars with $K_p<20$ that were observed by Kepler.  In the middle panels 
we limit the sample 
to include only stars with contamination $c<0.4$ to reduce the effect that stray 
light from nearby stars may have on the frequency of Kepler candidates, which may be significant in clusters like 
NGC~6791.  This value of $0.4$ is chosen fairly arbitrarily, and a more thorough analysis is needed to determine 
the proper cutoff in $c$ to draw robust conclusions.  Finally, in the right panels, we limit the Kepler sample to only 
include stars that would lie between the single-star and equal-mass binary sequences of the respective clusters.  
We use \citet{2008A&A...482..883M} isochrones in log($T_{eff}$) vs. $K_p$ space to perform this selection.  

Although the vast majority of the Kepler observed stars (and especially those targeted for the ``EX" program) have 
$K_p < 16$, we include all stars with $K_p<20$ observed by Kepler to remain consistent with our theoretical 
expectation that planets have the highest detection probability between $16 < K_p < 20$ in NGC~6791.  
Given the old age and distance of NGC 6791, the cluster MS begins at $K_p \geq 16$ 
(Figure\ \ref{plot:hrd}).  Consequently, the majority of the observed Kepler candidates in Figure\ \ref{plot:nposition} 
for NGC~6791 must be giants, or possibly subgiants, if they are really around cluster members.  
Our numerical models suggest that it is unlikely for Kepler to detect more than about $3$ planets around 
giant host stars in NGC~6791 (Figure\ \ref{plot:n_kp}).  
Kepler has not yet reported strong planet candidates transiting stars that are within $1r_t$ from the center 
of NGC~6791 and have magnitudes and temperatures consistent with a MS cluster member 
(top-right panel of Figure\ \ref{plot:nposition}).  
Given that our models predict significantly more planet detections possible by Kepler for the MS 
stars in NGC 6791, we suggest that a detailed analysis of these fainter stars for planet transits would be highly valuable.   

If a transit is detected in the direction of a cluster, it is not a priori clear whether the planet is indeed around 
a cluster star or around a foreground/background star.  We find that if a transiting planet is detected within $0.5\,r_t$ 
from the center of NGC~6791 and the host star has properties such that it will lie within the single and equal-mass 
binary MSs in the CMD of NGC~6791, then the planet is very likely to be around a true cluster member.  
About $60\%$ of all transiting planet detections around stars satisfying this criteria and within $r_t$ from the center of 
NGC~6791 are expected to be around true cluster members (Figure\ \ref{plot:nposition}, top-right).  Given that there are 
about 10 Kepler candidates within the tidal radii of the 4 open clusters in the Kepler FOV 
(Figure\ \ref{plot:nposition}, bottom-right) around stars with properties satisfying the CMD based limits 
described above, about $6$ of these systems may be true cluster members.  The relative contributions 
from the cluster stars and the foreground/background stars do not change significantly if the power-law exponent for 
the planet-size distribution is changed in the range explored in this study (Table\ \ref{tab:np_rpdist}).  However, 
note that this part of the analysis has a few caveats as described below.       

We compute the predicted number of Kepler detectable planets transiting foreground/background stars 
superimposed on our model of NGC~6791 (thin lines in the top panels 
of Figure\ \ref{plot:nposition}).   To produce these curves, we embed the simulated cluster within a uniform field of 
background/foreground stars and treat these data in the same manner as for the Kepler observations.  
We use the Besan\c{c}on model of the Milky Way \citep{2003A&A...409..523R} to estimate the $K_p$-dependent 
surface densities of stars in the direction of NGC 6791.  Using the transformation equations 
from \citet{2002AJ....123.2121S} we convert the Johnson $B$, $V$ magnitudes given in the Besan\c{c}on model 
to SDSS $g$, $r$, and then use the standard Kepler conversions to find $K_p$.  
We estimate the surface densities of background/foreground stars in the 
direction of NGC~6791 as a function of $K_p$.  
The surface density of the expected Kepler detections for planets transiting these background/foreground 
stars per solid angle is then estimated using the same method described in Section\ \ref{sec:models} for $K_p$ values 
between $5<K_p<20$.  
These estimated surface densities are then summed to obtain the integrated surface density of planet detections 
for non-cluster stars over the full range in $5<K_p<20$.  
The expected cumulative number of transit detections for non cluster stars as a function of the distance ($r$) from the cluster center 
is then found simply by multiplying the expected number of transit detections per sky-projected area and $\pi r^2$.  

We find that the total number of background/foreground stars estimated from the Besan\c{c}on model 
can be lower by $\sim 30\%$ than that counted from the Kepler Input Catalog (KIC) by centering at the same Galactic 
latitude but away from the cluster.  Even the numbers of stars estimated from the KIC catalog using a center 
at different nearby locations at the same Galactic latitude can vary by about $10\%$.  These differences 
in the number of stars are directly translated to differences in the estimated number of detectable transits around 
non cluster members.  For example, using the KIC stars within an angular radius of $23.1^{'}$ centered around a point 
away from NGC~6791 but at the same Galactic latitude as NGC~6791, this model would estimate an expected number of detectable 
planets within $r < 1\,r_t$, transiting non cluster stars with $K_p<17$ to be $49$.  Using the same cut-offs for 
$K_p$ and $r$, but using the Besan\c{c}on model the estimated number of detectable transiting planets is $34$.  
We caution that, both of the above estimates are likely upper limits on the number of
detectable planets transiting non-cluster stars, since it assumes that
all non-cluster stars are dwarfs.
A significant fraction of the foreground/background stars will be
giant or subgiant stars for which the minimum detectable planet size
is significantly larger than for a MS star of the same
magnitude.
Thus, the thin lines should be treated as rough estimates for an upper
limit to the number of contaminating planets around
foreground/background stars.
Fortunately, even these results suggest that planet candidates within
$0.5\,r_t$ are likely cluster members if the selected sample has properties 
consistent with their being on the MS of the cluster.

A more complete and careful study of the frequency of planets in the fields near the open clusters 
in the Kepler survey is of great interest (but beyond the scope of this paper).  We stress here that a complete study 
of any putative trend between planet occurrence rate and position relative to a cluster 
must be mindful of a number of caveats (in addition to those mentioned above) 
related to how the sample of target stars 
are chosen before drawing secure conclusions.  For example, the search for planets in clusters is still 
incomplete (for NGC~6791 it has barely started) due to technical details related to targeting and the pipeline.  A careful study must include 
proper cuts in the $K_p$ values to fairly compare cluster stars with those in the field, and take into account 
differences in magnitude, stellar radii, contamination, pipeline selection effects between cluster and 
field stars, and preferably kinematic cluster membership information.  

Acknowledgmentes: We thank the referee, R.~Gilliland, for a detailed review and many helpful suggestions and 
information.  We thank S.~Meibom for helpful discussion.  We thank D.~Stello, A.~V.~Tutukov, K.~F.~Brogaard, 
for helpful suggestions and comments based on the preprint.  SC acknowledges support from
the Theory Postdoctoral Fellowship from UF Department of Astronomy and
College of Liberal Arts and Sciences.  This material is based on work supported by the National Aeronautics
and Space Administration under grant NNX08AR04G issued through the
Kepler Participating Scientist Program.  All cluster simulations using CMC were performed 
using the computer cluster fugu at Northwestern University.  F.A.R.\ acknowledges support from NASA Grant NNX12AI86G. 


\label{lastpage}
\end{document}